\documentclass[12pt]{article} 
\usepackage{latexsym, graphics, makeidx, epsfig, psfrag}

\renewcommand{\theequation}{\thesection.\arabic{equation}}
\newcommand{\be}{\begin{equation}}
\newcommand{\ee}{\end{equation}}
\newcommand{\bear}{\begin{eqnarray}}
\newcommand{\eear}{\end{eqnarray}}
\newcommand{\ba}{\begin{array}}
\newcommand{\ea}{\end{array}}
\newcommand{\lae}{\begin{array}{c}\,\sim\vspace{-21pt}\\<
\end{array}}
\newcommand{\gae}{\begin{array}{c}\,\sim\vspace{-20pt}\\>
\end{array}}

\newcommand{\mus}{m}
 
\newcommand{\CO}{{\cal O}} 
   
\begin{document}

\pagestyle{empty} 
\begin{titlepage}
\def\thepage {} 

\title{\Large \bf  Spin-Dependent Macroscopic Forces from \\ [3mm]
New Particle Exchange \\ [6mm] {} }

\author{\normalsize
\bf \hspace*{-.3cm}
Bogdan A.~Dobrescu$\,^1$, Irina Mocioiu$\,^{2}$\\ \\
{\small {\it
$^1$ Theoretical Physics Department, Fermilab, Batavia, IL 60510, USA
}}\\
{\small \tt bdob@fnal.gov}\\ [2mm]
{\small {\it$^2$ Pennsylvania State University,
University Park, PA 16802, USA}}\\
{\small \tt irina@phys.psu.edu} \\  [2mm] {}  }

\date{ \normalsize May 31, 2006}
\maketitle
\vspace*{-12.4cm}
\noindent 
\makebox[11cm][l]{\small hep-ph/0605342} \
\makebox[11cm][l]{\small FERMILAB-Pub-06-084-T} \\

\vspace*{12.8cm}

\begin{abstract} 
\baselineskip=20pt 
{\normalsize 
Long-range forces between macroscopic objects are mediated by
light particles that interact with the electrons or
nucleons, and include spin-dependent static components
as well as spin- and velocity-dependent components.
We parametrize the long-range potential between two fermions
assuming rotational invariance, and find 16 different
components. Applying this result to electrically neutral objects, 
we show that the macroscopic potential depends on 72 measurable
parameters.
We then derive the potential induced by the exchange of a new  
gauge boson or spinless particle, and compare the limits set by measurements
of macroscopic forces to the astrophysical limits on the couplings
of these particles. 
\\ }

\end{abstract}

\vfill\end{titlepage}
\baselineskip=20pt \pagestyle{plain} \setcounter{page}{1}

\tableofcontents 

\vspace*{.8cm}

\section{Introduction} \setcounter{equation}{0}

The electromagnetic and gravitational interactions, mediated by spin-1 and  
spin-2 particles, are the only macroscopic forces observed so far.
However, other  macroscopic forces could exist, and more sensitive 
measurements might reveal them. 
Searches for long-range spin-independent forces have a long history 
of substantial improvements achieved by various groups (for recent reviews
see Ref.~\cite{Adelberger:2003zx, Long:2003ta}). By contrast,
long-range spin-{\it dependent} forces could lead to a
broader variety of observable effects, but so far they have been less 
intensely investigated.
Most searches have been concentrated on two types of spin-dependent 
long-range forces that could be induced by axion exchange, the so-called
dipole-dipole and monopole-dipole interactions \cite{MoodyWilczek}. 

Measurements of forces between macroscopic polarized objects
have set limits on new
dipole-dipole potentials among electrons 
\cite{Ritter1990, Wineland1991, Pan:1992yr, Ni:1993ct, Ni1994},
and between electrons and nucleons \cite{Wineland1991, Ni:1993ct}.
There are also limits on monopole-dipole forces between polarized 
electrons and unpolarized objects 
\cite{Wineland1991, Youdin:1996dk, Ni:1999di, Heckel:1999sy}, 
as well as between polarized nucleons and unpolarized objects 
\cite{Wineland1991, Youdin:1996dk, Venema1992, Daniels:1991ec}.
Earlier experiments are reviewed in 
\cite{Wineland1991, Ni:1993ct, Adelberger:1992ph, Newman1994}.

Here we study spin-dependent forces 
between macroscopic objects that could exist given general 
assumptions within quantum field theory.
We focus on rotational-invariant potentials  that could be induced
by the exchange of new light particles, showing that several new 
kinds of spin-dependent macroscopic forces may exist and should
be searched for in experiments. 

The discovery of a new force with a range longer than about a
micrometer  would have a tremendous impact on our understanding of nature.
Furthermore, even if new macroscopic forces will not be discovered, 
setting limits on the various potentials is important for constraining 
many extensions of the Standard Model of particle physics.
The spontaneous breaking of continuous symmetries leads to the existence
of massless or very light (pseudo) Nambu-Goldstone bosons, such as axions,
familons, majorons, etc. \cite{Eidelman:2004wy}. It is also possible that 
new massless gauge bosons associated with unbroken gauge symmetries exist 
\cite{paraphoton}.
Such particles have naturally suppressed interactions with 
ordinary matter, but nevertheless could mediate long-range forces 
that may be accessible to laboratory experiments. 
As an application, we derive the 
limits on the couplings of a new massless spin-1 particle 
(``paraphoton'') from existing measurements of spin-dependent forces.

Massive spin-1 particles with general couplings, or bosons of 
spin-2 or higher, could also be light enough to mediate macroscopic forces,
albeit their low mass and feeble interactions would require very small 
dimensionless parameters or fine tuning.
We will show that the majority of the rotational-invariant spin-dependent  
potentials are generated by the exchange of a massive spin-1
particle in a Lorentz-invariant theory.

We first construct the most general 
momentum-space elastic-scattering  amplitude for two fermions consistent with
rotational invariance (see Section 2).
We then Fourier transform to position space in Section 3, and obtain the 
spin-dependent potential between two fermions.
In Section 4 we discuss the potential between macroscopic objects in the 
case of one-boson exchange in a Lorentz invariant theory (Section 4.1), 
as well as in more exotic cases, such as the exchange of a boson obeying 
a Lorentz-violating dispersion relation \cite{Arkani-Hamed:2004ar},
or the exchange of two or more particles (see Section 4.2).

We apply this general formalism to the case of spin-1 and spin-0
particle exchange in Sections 5 and 6, respectively. In this context
we compare the current experimental limits on spin-dependent forces
with the astrophysical limits on very light particles. 
Our results are summarized in Section 7.

\section{Long-range fermion-fermion interactions in momentum space}
\label{momentumspace}
\setcounter{equation}{0}

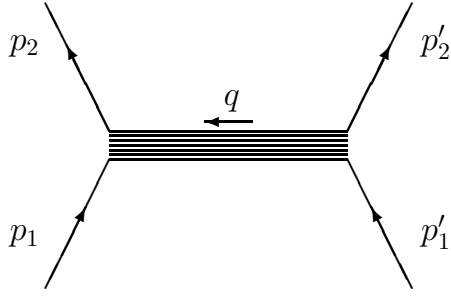
\begin{figure}[t]
\begin{center}
\scalebox{0.9}{
\begin{picture}(50,50)(20,20)
\thicklines
\put(0,6){\line(-1,2){27}}\put(0,6){\vector(-1,2){18}}
\put(0,-6){\line(-1,-2){27}}\put(-20,-46){\vector(1,2){10}}
\put(0,6){\line(1,0){100}}
\put(0,4){\line(1,0){100}}
\put(0,2){\line(1,0){100}}
\put(0,0){\line(1,0){100}}
\put(0,-2){\line(1,0){100}}
\put(0,-4){\line(1,0){100}}
\put(0,-6){\line(1,0){100}}
\put(100,6){\line(1,2){27}}\put(100,6){\vector(1,2){18}}
\put(100,-6){\line(1,-2){27}}\put(120,-46){\vector(-1,2){10}}
\put(60,10){\vector(-1,0){20}}\put(48,17){\large $q$}
\put(-42,40){\large $p_2$}
\put(-42,-40){\large $p_1$}
\put(131,40){\large $p_2^\prime$}
\put(131,-40){\large $p_1^\prime$}
\end{picture}
}
\end{center}
\vspace*{6em}
\caption{Elastic scattering of two fermions 
mediated by some very light particles represented generically by the 
horizontal blob of four-momentum $q$.
}
\label{fig:exchange}
\end{figure}

In order to derive the long-range force 
between two fermions of masses $m$ and $m^\prime$,
mediated by some very light particles, one needs to 
compute first the nonrelativistic limit of the scattering amplitude 
represented by the diagram shown in Figure \ref{fig:exchange}.
This amplitude can be expressed in terms of scalar invariants
formed out of the incoming and outgoing fermion three-momenta, 
$\vec p_1, {\vec p}_1^{\; \prime}$ and $\vec p_2, {\vec p}_2^{\; \prime}$, respectively, 
and the two fermion spins $\vec \sigma$ and $\vec \sigma\,'$.
In the center-of-mass frame only two momenta are independent, and we choose
the following linear combinations:
\bear
\vec q & \equiv & \vec p_2 - \vec p_1 \nonumber \\ [2mm]
\vec P & \equiv &  \frac{1}{2} \left(  \vec p_1 + \vec p_2  \right) ~.
\eear
Note that $\vec{q}$ is the momentum transferred to the fermion of mass $m$,
and $\vec{P}$ is the average momentum of that fermion.

With two spins and two momenta, one can construct 16 independent scalars
that include all possible spin configurations.
Eight of those include an even number of momenta, so they
are invariant under a parity transformation:
\bear
{\cal O}_1&=&1 ~~,
\nonumber \\[2mm]
{\cal O}_2&=&{\vec \sigma}\cdot\vec{\sigma}^{\,\prime} ~~,
\nonumber \\ [2mm]
{\cal O}_3&=& \frac{1}{\mus^2}\left(\vec \sigma \cdot \vec q\right)\,
\left( \vec{\sigma}^{\,\prime} \cdot \vec q\right) ~~, 
\nonumber \\[2mm]
{\cal O}_{4,5}&=&\frac{i}{2\mus^2} \left(\vec \sigma \pm \vec{\sigma}^{\,\prime} \right) 
\cdot \left(\vec P\times \vec q\right)  ~~,
\nonumber \\[2mm]
{\cal O}_{6,7}&=&\frac{i}{2 \mus^2}\left[
\left(\vec\sigma\cdot{\vec P}\right)\,
\left( \vec{\sigma}^{\,\prime} \cdot{\vec q}\right)\pm 
\left(\vec{\sigma}\cdot{\vec q}\right)\, 
\left( \vec{\sigma}^{\,\prime} \cdot{\vec P} \right)
\right]  ~~,
\nonumber \\[2mm]
{\cal O}_8 &=&  \frac{1}{\mus^2}
\left(\vec\sigma\cdot{\vec P}\right)\,
\left( \vec{\sigma}^{\,\prime} \cdot{\vec P}\right) ~~.
\label{ops}
\eear
We have included powers of the fermion mass $\mus$
in the denominators such that all these operators are dimensionless
(we use the natural unit system: $\hbar = c = 1$).
The other eight scalars change sign under a parity transformation:
\bear
{\cal O}_{9,10}&=& \frac{i}{2\mus}\, 
\left(\vec\sigma\pm \vec{\sigma}^{\,\prime}\right)\cdot {\vec q} ~~,
\nonumber \\[2mm]
{\cal O}_{11}&=&\frac{i}{\mus} \, 
\left(\vec\sigma\times \vec{\sigma}^{\,\prime}\right) \cdot {\vec q} ~~,
 \nonumber \\[2mm]
{\cal O}_{12,13}&=& \frac{1}{2\mus}\,  
\left(\vec\sigma\pm \vec{\sigma}^{\,\prime}\right) \cdot {\vec P} ~~,
 \nonumber \\[2mm]
{\cal O}_{14}&=&\frac{1}{\mus} \, 
\left(\vec\sigma\times \vec{\sigma}^{\,\prime}\right) \cdot {\vec P} ~~,
  \nonumber \\[2mm]
{\cal O}_{15}&=&\frac{1}{2\mus^3} \left\{\rule{0mm}{5mm}
\left[ \vec\sigma \cdot \left({\vec P}\times {\vec q} \right) \right]
\left( \vec{\sigma}^{\,\prime} \cdot \vec q \right) 
+ \left(\vec\sigma \cdot \vec q   \right) 
\left[\vec{\sigma}^{\,\prime}  \cdot \left(\vec P \times \vec q \right)  
\right]\right\}
\nonumber \\[2mm]
{\cal O}_{16}&=&\frac{i}{2\mus^3} \left\{\rule{0mm}{5mm}
\left[ \vec\sigma \cdot \left({\vec P}\times {\vec q} \right) \right]
\left( \vec{\sigma}^{\,\prime} \cdot \vec P \right) 
+ \left(\vec\sigma \cdot \vec P   \right) 
\left[\vec{\sigma}^{\,\prime}  \cdot \left(\vec P \times \vec q \right)  
\right]\right\} ~.
\label{ops-prime}
\eear
Any other scalar operator involving at least one of the two spins 
can be expressed as a linear combination of  the 
operators  ${\cal O}_i(\vec{q},\vec{P})$, $i = 1,\ldots,16$,
with coefficients that may depend on the momenta only through the
${\vec q\,}^2$ or $\vec{P}^2$ scalars. Note that energy-momentum
conservation implies ${\vec q}\cdot {\vec P} = 0$.
Examples of other operators which can be expressed as
linear combinations of ${\cal O}_i$, $i = 1,\ldots,16$,  
can be found in Appendix A. Although several of the operators given
in Eq.~(\ref{ops}) have been analyzed in the context 
of nuclear interactions \cite{Ericson:1988gk, Nagels:1975fb, Maessen:1989sx}, 
we believe that the complete set of 16 rotationally invariant 
operators has not been previously presented in the literature.

The amplitude for elastic scattering 
of the two fermions depends on the properties of the light 
particles that mediate it. 
The long-range nature of the force is due to the propagator of the exchanged 
particles, which is a function of the square of the four-momentum transferred, $q^2$.
Notice that $q_0 = 0$ due to energy conservation, so that $q^2 = - \vec{q\,}^2$. 
We use ${\cal P}(\vec{q\,}^2, m_0)$ to 
denote the imaginary part of the propagator with the Lorentz structure factored out.
The mass dimension of ${\cal P}(\vec{q\,}^2)$ is $-2$.
In the most common case, where the potential is induced by 
the exchange of one boson within a Lorentz invariant quantum field theory,
\be
{\cal P}(\vec{q\,}^2) = - \frac{1}{\vec{q\,}^2 + m_0^2} ~,
\label{standard-propagator}
\ee
where $m_0$ is the mass of the boson.
Other forms for the propagator are possible. For example,
in the case where {\it two} massless fermions are exchanged, 
the effective propagator takes the form \cite{Hsu:1992tg}
\be
{\cal P}(\vec{q\,}^2) = - \frac{1}{12\pi^2 M^2} 
\ln \left(\frac{\vec{q\,}^2}{M^2}\right) ~,
\label{twofermions}
\ee
where $M$ is the mass scale that suppresses the four-fermion 
contact interaction.
If Lorentz symmetry is violated, then a boson may have 
a kinetic term with four or more spatial derivatives, giving 
a propagator
\be
{\cal P}(\vec{q\,}^2) = - \frac{M^{2k-2}}{(\vec{q\,}^2)^k} ~,
\label{nonstandard-propagator}
\ee
where $k\ge 2$ is an integer, and $M$ is some mass scale.
The case $k=2$ has been studied in \cite{Arkani-Hamed:2004ar}.
For the moment we allow a generic form for ${\cal P}(\vec{q\,}^2)$,
assuming only that it leads to long-range forces.

Certain generic features of the
amplitude can be derived on general grounds.
The nonrelativistic amplitude between two fermions
may be written in the momentum space as
\be
{\cal A}\left(\vec{q},\vec{P}\right) =  {\cal P}(\vec{q\,}^2) 
\, \sum_{i=1}^{16}
{\cal O}_i\!\left(\vec{q},\vec{P}\right)  \, 
 f_i\!\left({\vec q\,}^2\!/m^2,\,\vec{P}^2\!/m^2 \right) ~,
\label{pot}
\ee
where $f_i$ are dimensionless scalar functions.
In the nonrelativistic limit, $f_i$ are polynomials 
with coefficients that depend on the 
couplings of the exchanged particles.
This is a general result based only on the assumption of rotational  
invariance (this assumption is not valid  
in certain Lorentz-violating field theories \cite{Arkani-Hamed:2004ar,Bluhm:1999ev}).

The physical interpretation of the 16 operators is more transparent in the 
position space, as discussed in the next section.

\section{Long-range potentials between fermions}
\label{sec:position}
\setcounter{equation}{0}

The Fourier transform of the momentum-space amplitude 
with respect to the momentum transfer $\vec{q}$ 
gives the position-space potential:
\be
V\!\left(\vec{r},\vec{v}\right) = - \int \frac{d^3q}{(2\pi)^3} e^{i\vec q\cdot \vec r}
\, {\cal A}\!\left(\vec{q}, m\vec{v}\,\right) ~,
\label{positionv}
\ee
where $\vec r$ is the position vector of the fermion of mass $m$ and initial momentum 
${\vec p}_1$ with respect to the fermion of mass $m^\prime$ and initial momentum 
$\vec{p}_1^{\,\prime}$. Note that in general the potential depends not only on the 
position $\vec{r}$, but also on the average velocity of the fermion of mass $m$
in the center-of-mass frame:
\be
\vec{v} = \frac{\vec{P}}{m} ~.
\ee

The inverse mass of the boson sets the range of the interaction, 
so that an experimental setup characterized by a distance scale
$r_{\rm exp}$ is sensitive to $1/m_0 \gae r_{\rm exp}$. 
We assume that $r_{\rm exp}$ is macroscopic, $r_{\rm exp} \gae O(1 \; {\rm mm})$,
The most important contributions to the potential come
from the momentum-independent terms of the
$f_i\!\left({\vec q\,}^2\!/m^2,\,\vec{v\,}^2 \right)$ polynomials in 
Eq.~(\ref{pot}). 
Additional powers of  $\vec{q\,}^2/m^2$ lead to terms of order
$\epsilon^2$ in the potential, where $\epsilon$ is of the order of 
$m_0/m$ or $1/(r_{\rm exp} m)$ (see Appendix B). 
Given that $m$ is the mass of the electron or nucleon, we find
$\epsilon < 10^{-10}$, so that it is a good approximation to  
include only the $\vec{q\,}^2=0$ pieces of the polynomials, 
$f_i\!\left(0,{\vec v\,}^2\right)$. 
Note that additional powers of  $\vec{q\,}^2/m^2$ also lead 
to Fourier transforms 
of the type $\delta\!\left(\vec{r}\right)$, or more singular ones,
which describe contact interactions rather than long-range 
potentials.

It is useful to observe that compared to ${\cal O}_1$ and
${\cal O}_2$, the operators ${\cal O}_i$ with $i =  9, 10, 11$ 
have effects of order $\epsilon$, the operators ${\cal O}_i$ with 
$i = 12, 13, 14$ have effects of 
order $v = |\vec v|$, while the remaining ones have effects suppressed by 
more powers of $\epsilon$ or $v$.
If the ${\vec q\,}^2$-independent term of $f_i$ vanishes, then 
the $\vec{q\,}^2/m^2$ term dominates, and for $i =  1,2, 9,\ldots, 14$
it might lead to experimentally observable effects.
By contrast, the operators ${\cal O}_i$ with  $i = 3,\ldots, 8, 15,16 $
are already quite suppressed, so that the $\vec{q\,}^2$-dependent terms 
of $f_i$ can be safely neglected in their case.
In what follows we will ignore the ${\vec q\,}^2$-dependent terms of 
all $f_i$, and we only mention that
if a physical situation would require the inclusion of some of them, then 
they could be treated similarly to the ${\vec q\,}^2$-independent 
terms.

The long-range potential between two fermions induced by a 
Lorentz-invariant, one-boson exchange 
can be written as 
\be
V\!\left(\vec{r},\vec{v}\right)= \sum_{i=1}^{16} 
{\cal V}_i\left(\vec{r}, \vec{v}\right) f_{i}\!\left(0,\vec{v\,}^2\right)~,
\label{potr}
\ee
where we defined a complete set of spin-dependent potentials, 
\be
{\cal V}_i\!\left(\vec{r},\vec{v}\right) 
= - \int \frac{d^3q}{(2\pi)^3} e^{i\vec q\cdot \vec r}
\, {\cal P}(\vec{q\,}^2)  \, 
{\cal O}_i\!\left(\vec{q},m\vec{v}\right)  ~,
\label{position-potentials}
\ee
with $i=1,\ldots, 16$.  As stated before, 
$f_{i}(0,\vec{v\,}^2)$ are polynomials in ${\vec v\,}^2$,
with coefficients given by dimensionless parameters 
that depend on the boson couplings to the fermions.

It is convenient to write the spin-dependent potentials in terms
of a dimensionless function of $r$:
\bear
y(r) & \equiv & - r \int \frac{d^3q}{(2\pi)^3} e^{i\vec q\cdot \vec r}
\, {\cal P}(\vec{q\,}^2) 
\nonumber\\ [3mm]
&=& -\frac{1}{2\pi^2} \int_0^\infty d |\vec q| \, {\cal P}(\vec{q\,}^2) 
\, |\vec q|  \sin\left(|\vec q| r\right) ~.
\label{y-function}
\eear
Using the operators ${\cal O}_i$ with $i=1, \ldots, 8$,
defined in Eq.~(\ref{ops}),
we obtain the following   long-range, parity-invariant potentials:
\bear
{\cal V}_{1} &=&\frac{1}{r}\, y(r) ~~,
\nonumber\\[3mm]
{\cal V}_{2} &=&\frac{1}{r} \;{\vec \sigma}\cdot {\vec \sigma\,'} \, 
y(r)~~,
\nonumber\\[3mm]
{\cal V}_{3} & = & \frac{1}{ m^2\,r^3} \left[ 
{\vec \sigma}\cdot {\vec \sigma\,'} \left(1 - r\frac{d}{dr} \right)
- 3 \left( \vec\sigma\cdot \hat{\vec{r}} \right)\,
\left({\vec \sigma\,'}\cdot \hat{\vec{r}} \right) 
\left(1 - r\frac{d}{dr} + \frac{1}{3}r^2\frac{d^2}{dr^2} \right) \right] y(r)~~,
\nonumber\\[3mm] 
{\cal V}_{4,5}  & = &-\frac{1}{2 m\,r^2}
\left(\vec \sigma \pm \vec{\sigma}^{\,\prime} \right) 
\cdot \left(\vec v\times \hat{\vec{r}} \right)  
\left(1 - r\frac{d}{dr}\right) y(r) ~~,
\nonumber \\[3mm] 
{\cal V}_{6,7} & = & -\frac{1}{ 2 m\,r^2}
\left[\rule{0mm}{5mm} \left(\vec\sigma\cdot{\vec v}\right)\,
\left( \vec{\sigma}^{\,\prime} \cdot \hat{\vec{r}} \right) 
\pm \left(\vec\sigma\cdot  \hat{\vec{r}} \right)\, 
\left( \vec{\sigma}^{\,\prime} \cdot{\vec v} \right) \right] 
\left(1 - r\frac{d}{dr}\right) y(r) ~~,
\nonumber \\[3mm]
{\cal V}_8 & = & \frac{1}{r} \left(\vec\sigma\cdot\vec v\right)
\left(\vec{\sigma\,}^\prime\cdot\vec v\right) \, y(r)  ~~, \;\;
\label{even}
\eear
where $r$ is the length of the $\vec r$ vector, and 
we have defined the unit vector
\be
\hat{\vec{r}} \equiv \frac{\vec r}{r} ~.
\ee   
The operators ${\cal O}_i$ with $i=9, \ldots, 16$,
defined in Eq.~(\ref{ops-prime}),
give rise to the following long-range, parity-violating potentials:
\bear
{\cal V}_{9,10} &=&-\frac{1}{2 m \, r^2}\, 
 \left(\vec\sigma\pm\vec\sigma^{\,\prime}\right) \cdot \hat{\vec{r}} \,
\left(1 - r\frac{d}{dr}\right) y(r)~~,  
\nonumber\\[3mm]
{\cal V}_{11} &=&-\frac{1}{ m \, r^2} \, 
\left(\vec\sigma\times \vec{\sigma}^{\,\prime}\right)\cdot \hat{\vec{r}} \,
\left(1 - r\frac{d}{dr}\right) y(r) ~~,
\nonumber\\[3mm] 
{\cal V}_{12,13} &=&
\frac{1}{2 r}\, \left(\vec\sigma\pm \vec{\sigma}^{\,\prime}\right) \cdot {\vec v} 
\; y(r)~~,
 \nonumber \\[3mm]
{\cal V}_{14} &=&\frac{1}{r} \, 
\left(\vec\sigma\times \vec{\sigma}^{\,\prime}\right) \cdot {\vec v} 
\; y(r)~~,
\nonumber\\[3mm]
{\cal V}_{15} &=&-\frac{3}{2 m^2 \, r^3} 
\left\{ \rule{0mm}{5mm}\left[ 
\vec\sigma\cdot \left(\vec v\times \hat{\vec{r}} \right) \right]
\, \left(\vec\sigma^{\,\prime} \cdot \hat{\vec{r}} \right)
+ \left(\vec\sigma \cdot \hat{\vec{r}} \right) \, 
\left[ \vec\sigma^{\,\prime} \cdot \left(\vec v\times \hat{\vec{r}}
\right) \right]\right\} 
\nonumber\\[2mm]
&& \times \left(1 - r\frac{d}{dr} + \frac{1}{3}r^2\frac{d^2}{dr^2}\right) y(r)~~,
\nonumber\\[2mm]
{\cal V}_{16} &=&   -\frac{1}{2m \, r^2} 
\left\{ \rule{0mm}{5mm}
\left[ \vec\sigma\cdot \left(\vec v\times \hat{\vec{r}} \right) \right]
\, \left(\vec\sigma^{\,\prime} \cdot \vec{v} \right)
+ \left(\vec\sigma \cdot \vec{v} \right) \, 
\left[ \vec\sigma^{\,\prime} \cdot \left(\vec v\times \hat{\vec{r}}
\right) \right]\right\} \left(1 - r\frac{d}{dr}\right) y(r) ~~.
\nonumber \\
\label{odd}
\eear
It is interesting that there are both parity-even ($i=2,3,6,7,8$)
and parity-odd ($i=11,14,15,16$) potentials
which induce macroscopic forces between two polarized objects.
Among those, ${\cal V}_{3}$ is the so-called dipole-dipole
potential.
Likewise, there are both  parity-even ($i=4,5$)
and parity-odd ($i=9,10,12,13$) potentials
which induce forces between one polarized and one unpolarized object.
The so-called monopole-dipole  potential is given by 
${\cal V}_9 + {\cal V}_{10}$.

Notice that in the case of identical fermions, only one linear combination
of the ${\cal V}_{4}$ and ${\cal V}_{5}$ potentials is relevant.
The same is true for the following pairs:  ${\cal V}_{6}$ and ${\cal V}_{7}$,
${\cal V}_{9}$ and ${\cal V}_{10}$, ${\cal V}_{12}$ and ${\cal V}_{13}$.

There are several static spin-dependent types of
long-range potentials: ${\cal V}_{2}, {\cal V}_{3}$, 
${\cal V}_9$, ${\cal V}_{10}$ and ${\cal V}_{11}$.
The other  potentials depend on the relative velocity of the two fermions.
In general, each of these potentials has an arbitrary coefficient that needs 
to be measured. 
Note though, that in simple models only some of the 16 potentials listed 
above are present.
In Sections 5 and 6 we will derive all the spin-dependent potentials 
that can arise in Lorentz-invariant quantum field theories from exchange 
of a spin-0 or spin-1 boson.

\section{Interactions between macroscopic objects}
\setcounter{equation}{0}

${\cal V}_i$ with $i=1, \ldots, 16$, given in Eqs.~(\ref{even}) and
(\ref{odd}), form a complete set of 
spin-dependent potentials between two fermions, assuming that 
rotational invariance is an exact symmetry of the Lagrangian.
To a good approximation, macroscopic objects are formed of electrons,
neutrons and protons, so that a sum over 
the potential between pairs of fermions belonging to two different objects
gives the total potential between those objects.
One should keep in mind though that this is just an approximation: 
some of the mass (a fraction of a percent) of a macroscopic 
object is due to the nuclear binding energy, which means that if there 
are long-range 
forces between electrons and gluons, for example, then their effects 
would not be fully taken into account by summing over fermion pairs.

In section 2 we have argued that the propagator of the very light
particles that mediate macroscopic forces may have various forms.
In this section we first discuss the case of standard propagator,
given in Eq.~(\ref{standard-propagator}), and later in subsection 4.2 we 
consider other forms for the propagator, as in  Eqs.~(\ref{twofermions})
and (\ref{nonstandard-propagator}).

\subsection{Exchange of one boson with standard propagator}
\label{standard}

In the case of one-boson exchange forces within a Lorentz-invariant
quantum field theory, the propagator (\ref{standard-propagator})
leads to a simple form for 
the function $y(r)$ defined in Eq.~(\ref{y-function}):
\be
y(r) = \frac{1}{4\pi} \, e^{-m_0 r} ~,
\ee
where $m_0$ is the mass of the boson exchanged.
The ensuing spin-independent potential, ${\cal V}_{1}$, 
is then of the well-known Yukawa 
type, such that the static potential between two point-like, 
unpolarized objects is given by
\bear
V_{1}(r) & = & \left\{\rule{0mm}{5mm}
N_e N_e^\prime \left[ \rule{0mm}{4mm} f_{1}^{ee}(0,0) + f_{1}^{pp}(0,0) + 2 f_{1}^{ep}(0,0)\right]
+  N_ n N_n^\prime f_{1}^{nn}(0,0) \right. 
\nonumber \\ [3mm]
&& + \;  \left. \rule{0mm}{5mm} \left(N_e N_n^\prime + N_n N_e^\prime\right) 
\left[ \rule{0mm}{4mm} f_{1}^{en}(0,0) + f_{1}^{ep}(0,0) \right]
 \right\} \frac{1}{4\pi r} e^{-m_0 r}  ~~,
\label{yukawa-potential}
\eear
where $N_e, N_n$ ($N_e^\prime, N_n^\prime$) 
are the number of electrons and neutrons in the first (second) object,
respectively, and we assumed that the objects are electrically neutral.
The coefficients $f_{1}^{ee}(0,0)$, $f_{1}^{\cal N N}(0,0)$ 
and $f_{1}^{e \cal N}(0,0)$, with ${\cal N} = n$ or $p$, depend on the
couplings of the exchanged boson to the electrons and nucleons,
and can be derived as shown in Eq.~(\ref{pot})
by computing the amplitudes for elastic $ee$, $e{\cal N}$ and ${\cal N}{\cal N}$ scattering,
respectively.
The macroscopic forces between unpolarized objects 
induced by the static potential in Eq.~(\ref{yukawa-potential})
have been studied in great detail (see, {\it e.g.},
Ref.~\cite{Adelberger:2003zx, Long:2003ta}).

Let us study now the 
spin-dependent forces between a point-like object whose electron spins
are polarized on average along a unit vector $\vec\sigma$,  
and a point-like unpolarized object. The average potential 
between an electron from the polarized object having the spin 
along $\vec\sigma$
and a neutron from the unpolarized object is given by adding the
contributions from ${\cal V}_{i}$ with $i=4,5,9,10,12,13$:
\be 
V_\sigma^{en} (\vec{r}, \vec{v}) = 
\frac{1}{8\pi r}\, \left\{ f^{en}_v \, \vec\sigma \cdot {\vec v} 
+ \left[ f^{en}_r \, \vec\sigma \cdot \hat{\vec{r}}
+ f^{en}_\perp \, \vec{\sigma} \cdot \left(\vec v\times \hat{\vec{r}} \right)  \right]
\frac{1+m_0 r}{m_e r} \right\} e^{-m_0 r}~~,
\label{sigma}
\ee
where $f^{en}_v$, $f^{en}_r$ and $f^{en}_\perp$ are 
the dimensionless coefficients of the potential when 
the electron spin is along 
the center-of-mass velocity $\vec v$ of the polarized object with respect to 
the unpolarized object,
along the unit vector $\hat{\vec{r}}$ pointing from the unpolarized object 
towards the polarized one, or along $\vec v\times \hat{\vec{r}}$, 
respectively.
These coefficients are given in terms of the polynomials 
$f_i$ introduced in Eq.(\ref{pot}) by 
\bear
f^{en}_v & = & f_{12}^{en}(0,0) + f_{13}^{en}(0,0) 
~~,
\nonumber\\[3mm]
f^{en}_r & = & - f_{9}^{en}(0,0) - f_{10}^{en}(0,0) 
~~,
\nonumber\\[3mm]
f^{en}_\perp & = & - f_{4}^{en}(0,0) - f_{5}^{en}(0,0) 
\label{sigmacoeff}
\eear
where the upper indices $e$ and $n$ indicate that the 
fermions of mass $m$ and $m^\prime$ discussed in general in 
sections 2 and 3 are now specified to be an electron and a neutron,
respectively.
We have included only the $\vec{q\,}^2$- and $\vec{P}^2$-  
independent terms in $f_i$ because the $\vec{q\,}^2$-dependent terms
give tiny corrections of order $(m_0/m_e)^2$ 
while $\vec{P}^2$-dependent terms give relativistic corrections
which are also negligible in experiments searching for new 
macroscopic forces.

The average potential 
between the electron spin and the protons or electrons in the unpolarized 
object, $V_\sigma^{ep}$ and $V_\sigma^{ee}$, respectively,
may be written analogously to Eq.(\ref{sigma}).
Then the total potential between the object containing the polarized electrons
and the unpolarized object is 
\be
V_{\sigma_e} (\vec{r}, \vec{v}) = N_e \sigma_e \left[
N_p^\prime\left( V_\sigma^{ee} + V_\sigma^{ep} \right) +
N_n^\prime V_\sigma^{en} \right] ~,
\label{e-sigma}
\ee
where $N_e$ is the total number of electrons in the 
polarized object, $\sigma_e$ is the  polarization (the average 
projection of the electron spins along $\vec\sigma$ in the 
polarized object), $N_p^\prime$ and $N_n^\prime$ are the numbers
of protons and neutrons in the unpolarized object. In writing 
the above equation we have assumed that the unpolarized object is 
electrically-neutral.
If the polarized object has the neutrons or protons polarized
instead of the electrons,
then the potentials $V_\sigma^{n} (\vec{r}, \vec{v})$
or $V_\sigma^{p} (\vec{r}, \vec{v})$ are given by
Eq.~(\ref{e-sigma}) with the index $e$ replaced appropriately
by $n$ or $p$ in Eqs.~(\ref{sigma})-(\ref{e-sigma}).
If the boson exchange induces in addition a spin-independent potential,
then the total potential is the sum of the
terms in Eq.~(\ref{e-sigma}) and (\ref{yukawa-potential}). 

In the case of two polarized objects there are 9 types of spin-spin
potentials. Three of those are static,
\bear
{\cal V}_{2} &=&\frac{1}{4\pi r}\;{\vec \sigma}\cdot {\vec \sigma\,'}\,
e^{-m_0 r}~~,
\nonumber\\[3mm]
{\cal V}_{3} & = & \frac{1}{4\pi m_e^2\,r^3} \left[ {\vec \sigma}\cdot {\vec
\sigma\,'} (1+m_0 r)- \left( \vec\sigma\cdot \hat{\vec{r}} \right)\,
\left({\vec \sigma\,'}\cdot \hat{\vec{r}} \right) (3+3 m_0 r+m_0^2 r^2)\right]  
\, e^{-m_0 r}~~,
\nonumber\\[3mm] 
{\cal V}_{11} &=&-\frac{1}{4\pi m_e \, r^2} \, 
\left(\vec\sigma\times \vec{\sigma}^{\,\prime}\right)\cdot \hat{\vec{r}}~ 
(1+m_0 r) \, e^{-m_0 r} ~~,
\label{static}
\eear
while the other six potentials depend on the relative velocity of the 
two objects:
\bear
{\cal V}_{6,7} & = & -\frac{1}{8 \pi m_e\,r^2}\left[
\left(\vec\sigma\cdot{\vec v}\right)\,
\left( \vec{\sigma}^{\,\prime} \cdot \hat{\vec{r}} \right) 
\pm \left(\vec\sigma\cdot  \hat{\vec{r}} \right)\, 
\left( \vec{\sigma}^{\,\prime} \cdot{\vec v} \right) \right] (1+m_0 r)\, e^{-m_0 r} ~~,
\nonumber \\[3mm] 
{\cal V}_8 & = & \frac{1}{4\pi r} \left(\vec\sigma\cdot\vec v\right)
\left(\vec{\sigma\,}^\prime\cdot\vec v\right) \, e^{-m_0 r}
  ~~, \;\;
\nonumber \\[3mm]
{\cal V}_{14} &=&\frac{1}{4\pi r} \, 
\left(\vec\sigma\times \vec{\sigma}^{\,\prime}\right) \cdot {\vec v}~ e^{-m_0 r} ~~,
\nonumber\\[3mm]
{\cal V}_{15} &=&-\frac{1}{8\pi m_e^2 \, r^3} 
\left\{ \rule{0mm}{5mm}
\left[ \vec\sigma\cdot \left(\vec v\times \hat{\vec{r}} \right) \right]
\, \left(\vec\sigma^{\,\prime} \cdot \hat{\vec{r}} \right)
+ \left(\vec\sigma \cdot \hat{\vec{r}} \right) \, 
\left[ \vec\sigma^{\,\prime} \cdot \left(\vec v\times \hat{\vec{r}}
\right) \right]\right\} \nonumber\\[2mm]
& &~~~~~~~~~~ \times  \left(3+3m_0r+m_0^2r^2\right) \, e^{-m_0 r} ~~,
\nonumber\\[3mm]
{\cal V}_{16} &=& -\frac{1}{8\pi m_e \, r^2} 
\left\{ \rule{0mm}{5mm}
\left[ \vec\sigma\cdot \left(\vec v\times \hat{\vec{r}} \right) \right]
\, \left(\vec\sigma^{\,\prime} \cdot \vec{v} \right)
+ \left(\vec\sigma \cdot \vec{v} \right) \, 
\left[ \vec\sigma^{\,\prime} \cdot \left(\vec v\times \hat{\vec{r}}
\right) \right]\right\} (1+m_0 r) \, e^{-m_0 r}  ~~. \nonumber \\ 
\label{velocity}
\eear
The total spin-spin potential between two macroscopic objects,
one of them containing $N_e$ polarized electrons with a
polarization $\sigma_e$ along $\vec\sigma$, and the other object containing 
$N_n^\prime$ polarized neutrons with a polarization $\sigma_n$ along 
$\vec\sigma^\prime$, is given by
\be 
V_{\sigma_e\sigma^\prime_n} (\vec{r}, \vec{v}) =
N_e\sigma_e N_n^\prime \sigma_n^\prime 
\sum_{i} f_i^{en}(0,0) \, {\cal V}_i (\vec{r}, \vec{v}) ~,\; 
\label{sigma-macro}
\ee
where the sum is over the potentials shown in Eqs.~(\ref{static})
and (\ref{velocity}).
An analogous potential exists for two objects containing 
polarized electrons, except that all $n$ indices are replaced 
by  $e$, and the $f_i^{en}(0,0)$ coefficients may be obtained by computing
the $ee\rightarrow ee$ amplitude. Similar statements apply to 
the $ep, pp, nn$ or $np$  spin-spin potentials. Notice that
several of the potentials in Eqs.~(\ref{static}) and (\ref{velocity})
include an inverse power of the electron mass, $m_e$, introduced 
to keep the $f_i$ functions dimensionless. In the case of the 
potentials between nucleons, $m_e$ is replaced by $m_n$ (or else the
$f_i$ functions need to be rescaled appropriately).

We briefly discuss the experimental limits on the 
coefficients  of the various potentials.  
Tests of the equivalence principle and of the inverse square law 
set limits on the Yukawa potential between unpolarized objects. 
Given that the tests involve macroscopic objects which are electrically neutral,
the boson couplings to the electron and proton are not constrained 
separately. Only their sum is constrained at roughly the same 
level as the neutron vector coupling.
Thus, the limits may be expressed in terms of the three combinations of $f_{1}$ coefficients 
that appear in Eq.~(\ref{yukawa-potential}):
\be 
\left| f_{1}^{nn}(0,0) \right| \; , \; 
\left| f_{1}^{en}(0,0) + f_{1}^{ep}(0,0)\right|   \; , \; 
\left| f_{1}^{ee}(0,0) + f_{1}^{pp}(0,0) + 2 f_{1}^{ep}(0,0)\right|
< 10^{-40} - 10^{-48}  ~,
\label{f1-limit}
\ee
where the weaker limit applies to $1/m_0$ of order 1 cm, 
while the stronger one applies to  $1/m_0 > 10^8$ m, the Earth-Moon distance
(see Figure 4 of \cite{Adelberger:2003zx}).

The most stringent limit on the dipole-dipole potential ${\cal V}_3$ between electrons
is set in Ref.~\cite{Ni1994} (see also \cite{Pan:1992yr,Ni:1993ct}, where the 
potential is explicitly written\footnote{The potential 
shown in Eq.~(1) of Ref.~\cite{Ni:1993ct} falls off as $1/r$ instead
of $1/r^3$. We believe that this is just a typo.}):
$1.2\pm 2.0 \times 10^{-14}$ times the magnetic interaction
of two electrons, for $1/m_0 \gae 10$ cm. At the $1 \sigma$ confidence level we then find
\be
- 0.8 \times 10^{-14} < \frac{f_{3}^{ee}(0,0)}{4\pi^2\alpha} < 3.2 \times 10^{-14} ~.
\label{f3-limit-ee}
\ee
Similarly, the limit on the dipole-dipole potential between 
an electron and a neutron \cite{Wineland1991} gives
\be
\frac{\left|f_{3}^{en}(0,0)\right|}{4\pi^2\alpha |\mu_n/\mu_N|} < 2.3 \times 10^{-11} ~,
\label{f3-limit-en}
\ee
for $1/m_0 \gae 1$~m. Here $|\mu_n/\mu_N| \approx 1.913$ is
the ratio of the neutron magnetic moment to the nuclear magneton.
The limits on the dipole-dipole potential between nucleons, 
or between an electron and a proton are weaker \cite{Wineland1991}.

The static spin-spin potential  ${\cal V}_2$ has not been 
experimentally searched for. However, the limits on the dipole-dipole potential 
${\cal V}_3$ provide an indirect constraint on ${\cal V}_2$. 
It is not clear how accurate would be the use of the best limits on ${\cal V}_3$, 
given in Ref.~\cite{Ni1994} and \cite{Wineland1991}, to constrain 
${\cal V}_2$, because ${\cal V}_3$ includes a 
$(\sigma \cdot \vec r) (\sigma^\prime \cdot \vec r)$ piece which is not 
present in ${\cal V}_2$.
By contrast, the limit set in 
Ref.~\cite{Ritter1990} explicitly applies to the $\sigma \cdot \sigma^\prime$ piece
of the  dipole-dipole potential between electrons. Given that ${\cal V}_2$
falls off as $1/r$ while ${\cal V}_3$ falls off as $1/r^3$, we estimate
\bear
\left|f_{2}^{ee}(0,0)\right| & \!\!\lae \!\!& 
\frac{4\pi^2\alpha}{m_e^2 r_{\rm exp}^2}  10^{-11} 
\nonumber \\ [3mm]
&\!\!\approx \!\!&  4\times 10^{-35} ~,
\label{f2-limit}
\eear
where $r_{\rm exp} \approx 10$ cm is the typical distance probed in the experimental setup
of Ref.~\cite{Ritter1990}.

The only other 
static spin-spin potential, ${\cal V}_{11}$, has also not been directly tested.
To the best of our knowledge, the velocity-dependent spin-spin 
potentials, with coefficients given in Eq.~(\ref{velocity}), have not been 
experimentally constrained yet.

The monopole-dipole types of interaction given by ${\cal V}_9+{\cal V}_{10}$ 
[see second term in Eq.~(\ref{sigma})]
have also been experimentally searched for \cite{Youdin:1996dk, Ni:1999di, Heckel:1999sy}.
The most stringent limits 
have been obtained very recently for the interaction
between an object with  polarized electrons and an unpolarized object in 
Ref.~\cite{Heckel2006}: 
\be
\left| f_r^{en}(0,0) \right| \; , \; \left| f_r^{ee}(0,0) +  f_r^{ep}(0,0)\right| 
\lae 10^{-30} - 10^{-36} ~~,
\label{fr-limit}
\ee 
where the weaker limit applies to distances of order 1 m, while the stronger 
limit is valid for distances above $10^{11}$ m (the Earth-Sun distance).
These limits represent improvements by at least two orders of magnitude over
the previous ones given in Refs.~\cite{Youdin:1996dk, Ni:1999di}.
The best limits on the monopole-dipole potential between 
an object with  polarized neutrons and an unpolarized object, set in 
Ref.~\cite{Venema1992}, give
\be
\left| f_r^{nn}(0,0) \right| \; , \; \left| f_r^{ne}(0,0) +  f_r^{np}(0,0)\right| 
\lae 10^{-27}-10^{-33} ~~,
\label{fr-limit-n}
\ee 
for $1/m_0$ in the  $1-10^6$ m range.

Velocity dependent potentials of the type ${\cal V}_4+{\cal V}_{5}$ and 
${\cal V}_{12}+{\cal V}_{13}$ have been tested for the first time \cite{Heckel2006}
while this paper was being written, and the preliminary 
limits for distances above $10^{11}$ m are
\bear
\left|f_\perp^{ en}\right| \; , \; \left|f_\perp^{ ee}+f_\perp^{ ep}\right| 
&\!\! \lae \!\!& 10^{-32}  ~~,
\nonumber \\ [3mm]
f_v^{ en} \; , \; f_v^{ ee}+f_v^{ ep} 
&\!\! \lae \!\!& 10^{-55} ~~, 
\label{fv-limit}
\eear
with $f_\perp$ and $f_v$ defined in Eq.~(\ref{sigmacoeff}).
We do not show a lower limit for $f_v$, because the central value 
obtained in Ref.~\cite{Heckel2006} differs from zero by almost $2\sigma$.
Note that the limit on  $f_v$ is stronger by many orders of magnitude than 
the limit on any other $f_i$ coefficient.

\subsection{Non-standard dispersion relations} 
\label{nonstandard}

So far we have considered long-range potentials induced by the exchange of 
a boson whose propagator has the usual pole structure, $1/(q^2 - m_0^2)$,
leading to the standard dispersion relation $E^2={\vec q\,}^2+m_0^2$~.
This form for the propagator follows from the assumptions that the 
kinetic term is Lorentz invariant and quadratic in derivatives. 
If the kinetic term involves higher derivatives, 
the propagator would include higher inverse powers of $q^2$, and would 
lead to new structures for the potentials. However, such 
kinetic terms lead to instabilities or  unitarity violation, 
so they may not be allowed in well-behaved physical theories.

The propagator (and therefore the dispersion relation) may be modified if 
Lorentz symmetry is 
broken, because then the kinetic terms may involve quartic or higher 
spatial derivatives while the time derivatives are quadratic, as required
in a well behaved theory.
For example, a dispersion relation of the type $E^2={\vec q\,}^4/M^2$
appears in Ref.~\cite{Arkani-Hamed:2004ar}, where Lorentz symmetry is spontaneously broken.
One could imagine a
larger class of propagators for a boson which involve higher powers of
$1/\vec{q\,}^2$. In the case of the propagator shown in 
Eq.~(\ref{nonstandard-propagator}), which is of the $(\vec q)^{-2k}$ type
with $k\ge 2$ integer,
the function $y(r)$ defined in Eq.~(\ref{y-function}) may be computed 
using a Fourier transform given in Eq.~(\ref{fourier-ghost}):
\be
y(r) = 
\frac{1}{4\pi}\frac{1}{[2 (k-1)]!}{r^{2k-2}}   ~.
\ee
The spin-dependent potentials are given by  Eqs.~(\ref{even}) and (\ref{odd}).
Note that the $r$-dependence is different than in the case of a normal
one-boson exchange analyzed in Section 4.1. 
For example, for $k=2$ (the case analyzed in Ref.~\cite{Arkani-Hamed:2004ar}),
the static spin-spin potential  falls off as $1/r$:
\be
V(\vec{r}) \sim - \frac{1}{r} \left[ 
\left( \vec\sigma \cdot \vec{\sigma}^{\, \prime}\right)
- \left( \vec\sigma \cdot  {\hat{\vec{r}}}
 \right) \,
\left(  \vec{\sigma}^{\, \prime} \cdot  {\hat{\vec{r}}}
\right) \right] ~.
\label{ghost}
\ee

Another case of interest is the long-range potential induced by exchange 
of two or more particles. A well known example is the force due to
two-neutrino exchange \cite{Hsu:1992tg,Feinberg:1989ps}. 
The one-loop diagrams involving two neutrinos are equivalent to 
the tree-level exchange of a single boson with an effective propagator 
of the type $\sim \ln\left(\vec{q\,}^2\right)$, as shown in Eq.~(\ref{twofermions}). 
The Fourier transform leads
to a potential which falls off as $1/r^5$, and includes a spin-independent
term as well as spin-spin terms.

The exchange of two bosons has also been shown to lead to
additional types of potentials \cite{Feinberg:1989ps}. In particular, 
spin-independent potentials falling off as $1/r^3$, $1/r^5$ or $1/r^7$ 
are induced by the exchange of two spin-0 particles 
\cite{Grifols:1994zz}. Unfortunately, the strength 
of any of the two-particle-exchange macroscopic forces studied so far 
is many orders of magnitude smaller than the current experimental 
sensitivity to new particles.

\section{Spin-1 exchange forces}
\setcounter{equation}{0}


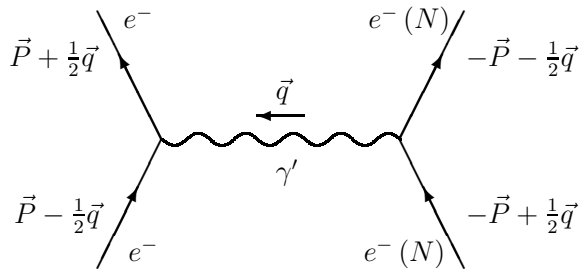
\begin{figure}[b]
\begin{center}
\scalebox{0.9}{
\begin{picture}(50,50)(20,20)
\thicklines
\put(0,0){\line(-1,2){27}}\put(0,0){\vector(-1,2){18}}
\put(0,0){\line(-1,-2){27}}\put(-20,-40){\vector(1,2){10}}
\multiput(0,0)(20,0){5}{\qbezier(0,0)(5,-5)(10,0)\qbezier(10,0)(15,5)(20,0)}
\put(100,0){\line(1,2){27}}\put(100,0){\vector(1,2){18}}
\put(100,0){\line(1,-2){27}}\put(120,-40){\vector(-1,2){10}}
\put(60,10){\vector(-1,0){20}}\put(48,17){$\vec q$}
\put(48,-17){$\gamma^\prime$}
\put(-64,31){$\vec{P} + \frac{1}{2}\vec{q}$} \put(-16,47){$e^-$}
\put(-62,-35){$\vec{P} - \frac{1}{2}\vec{q}$} \put(-14,-51){$e^-$}
\put(128,31){$- \vec{P} - \frac{1}{2}\vec{q}$}\put(87,47){$e^- \, (N)$}
\put(128,-35){$- \vec{P} + \frac{1}{2}\vec{q}$}\put(85,-51){$e^-\, (N)$}
\end{picture}
}
\end{center}
\vspace*{6em}
\caption{Paraphoton-exchange amplitude for nonrelativistic electron-electron or
electron-nucleon scattering. The three-momenta shown here correspond to the 
center-of-mass frame.}
\label{fig:paraphoton}
\end{figure}

The electromagnetic interaction is the only known long-range force
induced by a spin-1 particle. 
Nevertheless, low-mass spin-1 particles other than the photon may exist,
and they would lead to additional long-range forces that could be 
searched for in experiments.

\subsection{New massless gauge boson}

A spin-1 particle is naturally kept massless by an unbroken gauge symmetry.
In particular, a new $U(1)$ gauge symmetry would require the existence
of a massless spin-1 particle, labeled $\gamma^\prime$ and called 
paraphoton \cite{Holdom:1985ag}. 
If any of the Standard Model fields would be charged 
under the new $U(1)$ symmetry, then the gauge anomaly cancellation requires
the $U(1)$ charge to be proportional to the $B-L$ number, so that 
the $\gamma^\prime$ coupling to any electrically-neutral 
macroscopic object is proportional to the number of neutrons \cite{paraphoton}.
As a result, tests of the equivalence principle and of the inverse square law 
(see \cite{Blinnikov:1995kp} for a related discussion)
set an upper limit on the gauge coupling of $\gamma^\prime$ orders of magnitude 
below $10^{-19}$, which appears unnatural 
and also poses theoretical challenges \cite{Arkani-Hamed:2006dz}.

It is possible, however, that all Standard Model fields have zero
charge under the  new $U(1)$ symmetry, and yet $\gamma^\prime$ 
may interact with the quarks and leptons via dimension-6 operators 
involving two fermion fields, a paraphoton, and a Higgs doublet \cite{paraphoton}.  
Those operators are gauge invariant and do not depend on the fermion charges.
Replacing the Higgs doublet by its vacuum expectation 
value (VEV), $v_h \simeq 174$ GeV,  yields dimension-5 operators in the Lagrangian, 
representing magnetic- and electric-like dipole moments:
\be
\label{nucleon}
{\cal L}_{\gamma^\prime} =  \frac{v_h}{M^2} P_{\mu\nu} \left[ \,
\overline{e}\,\sigma^{\mu\nu}\! \left({\rm Re}C_e + i {\rm Im}C_e\gamma_5\right) e
\, + \, \sum_{{\cal N} = n,p} \overline{\cal N}\sigma^{\mu\nu}\! 
\left({\rm Re}C_{\cal N} + i {\rm Im}C_{\cal N} \, \gamma_5 \right) {\cal N} 
\right]~.
\ee
Here $P_{\mu\nu}$ is the field strength of the paraphoton, $e$ is the electron 
field, ${\cal N}$ is the nucleon field, while $C_e$ and $C_{\cal N}$ are dimensionless
complex parameters (their values are expected to be much less than unity). 
The  $\gamma^\prime$ coupling to nucleons is an effective low-energy 
Lagrangian that arises from a similar coupling of $\gamma^\prime$
to $u$ or $d$ quarks. These couplings may have different strengths, and therefore
the values of $C_{\cal N}$ when ${\cal N}$ is a proton or a neutron may be different.
The mass $M$ sets the scale where 
the dimension-6 operators are generated within an underlying  
theory which is well-behaved in the ultraviolet (examples of renormalizable models of 
this type are given in \cite{paraphoton}).

One $\gamma^\prime$ exchange between electrons or nucleons leads 
to a long-range force between chunks of ordinary matter.
In Figure \ref{fig:paraphoton} we show the three-momentum flow for the
scattering of fermions mediated by $\gamma^\prime$.
The amplitude for this process is given by 
\be
\label{p-amplitude}
{\cal A}\!\left(\vec{q},\vec{P}\right) = 
-\frac{1}{\vec{q\,}^2} \frac{4v_h^2}{M^4} \, S^\nu S^\prime_\nu ~,
\ee
where $\nu = 0,1,2,3$ is a Lorentz index, and we have defined 
\be
S^\nu = \overline{u}_e(P + q/2) \,
q_\mu\sigma^{\mu\nu}\!\left({\rm Re}C_e + i {\rm Im}C_e\gamma_5\right)\,
u_e(P - q/2) ~.
\label{s-nu}
\ee
The spinor $u_e(p)$ describes the electron field of four-momentum $p$. 
In the case of $e^-e^-$ scattering, $S^{\prime\nu}$ is identical to
$S^\nu$ except for the spinor $u_e(p^\prime)$ which depends on the 
momentum of the second electron.
In the case of $e^-{\cal N}$ scattering, $S^{\prime\nu}$ has the same structure 
as $S^\nu$ but the nucleon spinor $u_{\cal N}(p^\prime)$ and 
complex parameter $C_{\cal N}$ replace the electron ones.

In what follows we compute the nonrelativistic amplitude for 
$e^-{\cal N}$ scattering, because the result 
can be immediately adapted to $e^-e^-$ or ${\cal N}{\cal N}$ scattering.
In the nonrelativistic limit, the time-like component of $S^\nu$ is given by
\be
S^0 = - {\rm Im}C_e \, \vec{q}\cdot \vec{\sigma} + \frac{1}{m_e} {\rm Re}C_e
\left[ \left( \vec{P}\times \vec{q} \right) \cdot \vec{\sigma}
-\frac{i}{2} \vec{q\,}^2 \right] ~~.
\ee
Relativistic corrections to $S^0$, of order $\vec{P}^2/m_e^2$ 
and $\vec{q\,}^2/m_e^2$, do not introduce new spin-dependent terms. 
For the nucleon of initial three-momentum $-\vec{P}+\vec{q}/2$,
\be
S^{\prime 0} = {\rm Im}C_{\cal N} \, \vec{q}\cdot \vec{\sigma\,}^\prime 
+ \frac{1}{m_{\cal N}} {\rm Re}C_{\cal N}
\left[ \left( \vec{P}\times \vec{q} \right) \cdot \vec{\sigma\,}^\prime
-\frac{i}{2} \vec{q\,}^2 \right] ~~.
\ee

In order to compute the space-like components of $S^\nu$ and $S^{\prime \nu}$, 
it is useful to recall that energy-momentum conservation implies 
$\vec{P}\cdot \vec{q} = 0$ and $q^0 = 0$. We find
\bear
&& \vec{S} = {\rm Re}C_e \left[\vec{q}\times \vec{\sigma} 
- \frac{i\vec{q\,}^2}{4m_e^2} \vec{P} 
+ \frac{1}{m_e^2} \left(\vec{P} \cdot \vec{\sigma} \right) \vec{P}\times\vec{q} \,\right]
- \frac{1}{m_e}{\rm Im}C_e \, \left(\vec{q} \cdot \vec{\sigma} \right) \vec{P} ~~,
\nonumber \\ [3mm]
&& \vec{S\,}^\prime = - {\rm Re}C_{\cal N} \left[\vec{q}\times \vec{\sigma\,}^\prime 
- \frac{i\vec{q\,}^2}{4m_{\cal N}^2} \vec{P} 
+ \frac{1}{m_{\cal N}^2} \left(\vec{P} \cdot \vec{\sigma\,}^\prime \right) 
\vec{P}\times\vec{q} \,\right]
- \frac{1}{m_{\cal N}}{\rm Im}C_{\cal N} \, 
\left(\vec{q} \cdot \vec{\sigma\,}^\prime \right) \vec{P} ~~,\nonumber\\
\eear
with relativistic corrections affecting only the above spin-dependent
terms.

A lengthy but straightforward computation of the right-hand side of 
Eq.~(\ref{p-amplitude}) then gives the nonrelativistic amplitude. 
For the purpose of deriving the long-range potential, we may ignore 
the terms in $S^\nu S^\prime_\nu$ proportional to $\vec{q\,}^2$,
because upon Fourier transforming to position space they give only
contact interactions  or contributions additionally suppressed by $m_0$, as discussed in Section 3.1.
The result takes the form of Eq.~(\ref{pot}), with the functions
$f_i\!\left(0,\vec{v\,}^2 \right)$ being nonzero only for 
$i=3,15$. In the nonrelativistic limit, 
keeping only the leading order in $\vec{v\,}^2$, 
we obtain the following values for these functions:
\bear     
f_3^{e{\cal N}}(0,0) & = & - \frac{4v_h^2 m_e^2}{M^4} \; 
{\rm Re}\left( C_e^* C_{\cal N}\right) ~,
\nonumber \\ [2mm]
f_{15}^{e{\cal N}}(0,0) & = &  \frac{4v_h^2 m_e^2}{M^4} 
\left( 1 + \frac{m_e}{m_{\cal N}}\right)  
\left[ {\rm Re}\left( C_e\right) {\rm Im} \left( C_{\cal N}\right) -
{\rm Im}\left( C_e\right) {\rm Re} \left( C_{\cal N}\right)
\rule{0mm}{4mm}\right] ~~.
\label{paraphoton}
\eear
Therefore, the long-range potential between an electron and a nucleon 
induced by $\gamma^\prime$ exchange is given by
\be
V_{e{\cal N}}\left(\vec{r},\vec{v}\right)= \sum_{i=3,15} 
\left.{\cal V}_i\left(\vec{r},\vec{v}\right) \rule{0mm}{5mm}\right|_{ m=m_e} 
f_{i}^{e{\cal N}}\!\left(0,0\right)
\label{p-pot}
\ee
where the parity-even potential ${\cal V}_3$ 
is given in 
Eq.~(\ref{even}) while the parity-odd potential ${\cal V}_{15}$ 
is given in Eq.~(\ref{odd}). 
Notice that the long-range potential induced by paraphoton exchange may 
be observed only if both objects are polarized.

The long-range potential between electrons due to  $\gamma^\prime$ 
may be obtained from Eqs.~(\ref{p-pot}) and (\ref{paraphoton}) by
replacing the subscript ${\cal N}$ by $e$ (note that 
$f^{ee}_{15} = 0$, so that the long-range potential is static in this case):
\be
V_{ee}\left(\vec{r},\vec{v}\right)= -\frac{4v_h^2 m_e^2}{M^4} \;
|C_e|^2  \,\left. {\cal V}_3 
\rule{0mm}{5mm}\right|_{ m=m_e}  ~~.
\label{ee-p-pot}
\ee
The proton-proton and neutron-neutron long-range potentials have analogous forms 
with the appropriate replacements of $m_e$ and $C_e$ by the proton and 
neutron parameters. The  proton-neutron potential may include 
in addition the  ${\cal V}_{15}$ spin-dependent potential, similarly to 
Eq.~(\ref{p-pot}).
Note that the only static potential induced by $\gamma^\prime$ exchange is 
${\cal V}_3$, which gives the usual long-range force between two magnetic
dipole moments but with an overall strength that depends on the 
$\gamma^\prime$ couplings.

Given that the dimension-six operators that give rise to the 
effective $\gamma^\prime$ couplings in Eq.~(\ref{nucleon}) involve 
a chirality flip of the fermions, it is expected that its dimensionless
coefficients are of the order of or smaller than 
the corresponding Yukawa coupling to the Higgs doublet. 
It is therefore useful to factor out the Yukawa coupling from the 
$C_e$ and $C_{\cal N}$  parameters: 
\bear
c_e \equiv \frac{v_h}{m_e} |C_e| ~,
\nonumber \\ [3mm]
c_{\cal N} \equiv \frac{v_h}{m_d} |C_{\cal N}| ~,
\eear
where $m_d$ is the down-quark mass. The parameters 
$c_e$ and $c_{\cal N}$ may be as large as $O(1)$, but could be orders of magnitude 
smaller than one if the dimension-6 operators are generated at loop level in some
renormalizable model with weakly coupled fields.
The experimental limits on the paraphoton coupling to the electrons and
nucleons can be expressed in terms of 
$M/\sqrt{c_e}$ and $M/\sqrt{c_{\cal N}}$, respectively.

The static potential between electrons induced by $\gamma^\prime$  
exchange takes the form
\be
V(\vec r)= - \frac{c_e^2 m_e^2}{\pi M^4 r^3}  \left[\vec
\sigma_1\cdot\vec\sigma_2-
3 \left( \vec\sigma_1\cdot\hat{\vec r} \right)\, 
\left(\vec \sigma_2 \cdot \hat{\vec r}\right) \right] ~,
\ee
so that it is an attractive ${\cal V}_3$ potential. 
Using the  $1 \sigma$ limit shown in Eq.~(\ref{f3-limit-ee}),
we find
\be
\frac{c_e^2 m_e^4}{\pi^2 \alpha M^4 } < 0.8 \times 10^{-14} ~,
\label{v3-gamma}
\ee
where $\alpha$ is the fine structure constant.
This translates into a limit 
\be
\frac{M}{\sqrt{c_e}} > 3.3 \; {\rm GeV} ~.
\label{limit-ce}
\ee

Similarly, the limit on the dipole-dipole potential between 
an electron and a neutron \cite{Wineland1991}, shown in Eq.~(\ref{f3-limit-en}), 
gives
\be
\frac{c_e c_{\cal N} \,m_e^2 m_d m_p}{\pi^2 \alpha M^4 |\mu_n/\mu_N|} 
\cos\left( \theta_e - \theta_{\cal N} \right) < 2.3 \times 10^{-11} ~,
\ee
where $\theta_{e, {\cal N}}$ are the complex phases of 
$C_{e, \cal N}$ and $m_p$ is the proton mass.
We find the following constraint on the paraphoton couplings:
\be
\frac{M}{\sqrt{c_e c_{\cal N}}} 
\cos^{-1/4}\!\left(\theta_e -\theta_{\cal N}\right) > 
4.2 \; {\rm GeV} ~.
\label{limit-cecn}
\ee
where we used $m_d \sim 4$ MeV.

The bounds from star cooling  on $M/\sqrt{c_e}$ and $M/\sqrt{c_{\cal N}}$ 
are three orders of magnitude stronger than Eqs.~(\ref{limit-ce}) and 
(\ref{limit-cecn}), while the limits from
primordial nucleosynthesis are also substantially stronger
($M/\sqrt{c_e} \gae 100$ GeV and 
$M/\sqrt{c_{\cal N}} \gae 400$ GeV) \cite{paraphoton}.
However, there are potential loopholes in these astrophysical
and cosmological limits, whereas the limit from searches for new 
macroscopic forces is robust. A well known loophole in the 
limit from primordial nucleosynthesis is the possibility of 
a chemical potential during the early Universe. 
The limits from star cooling, considered unavoidable
in the case of axions \cite{Eidelman:2004wy}, 
could be avoided in the case of the 
paraphoton if the properties of this massless gauge boson 
depend on temperature. We contemplate a theory that besides the
Standard Model and the new $U(1)$ gauge group includes two 
or more scalar fields such that there is a mechanism of symmetry 
non-restoration at high temperatures \cite{Weinberg:1974hy}.
Specifically, if a scalar charged under the new $U(1)$
acquires a VEV when in thermal equilibrium 
in a star, and if this VEV is larger than the 
star temperature, then the $\gamma^\prime$  
emission from the star is exponentially suppressed.
As a result, star cooling via $\gamma^\prime$ emission could be 
negligible.

\subsection{General spin-1 exchange}

So far we have discussed the case of a massless spin-1 field which couples
to electrons or nucleons via higher-dimensional operators.
Let us turn now to a more general Lorentz-invariant extension of the Standard Model
that includes a new spin-1 field, $Z^\prime$, that is electrically neutral.
We assume that its mass $m_0$ is nonzero but smaller than 
$10^{-3}$ eV, so that $Z^\prime$ exchange mediates forces with a range longer than
a micrometer. 
We will consider in some cases a mass as small as $10^{-18}$ eV, which is the 
inverse Earth-Sun distance.

Without loss of generality, we assume that such a $Z^\prime$ field
is the gauge boson associated with a new $U(1)_z$ gauge symmetry
that is spontaneously broken by the VEV of a
spin-0 field $\varphi$, which is a singlet under the Standard Model
gauge group.
The $Z^\prime$ mass is then related to the gauge coupling $g_z$ and 
the $\varphi$ charge $z_\varphi$: 
\be
m_0 = z_\varphi g_z \langle \varphi \rangle ~.
\ee

The $Z^\prime$ boson couples to the leptons and quarks of the first generation 
as follows:
\be
{\cal L}_{Z'}^g= g_z Z'_\mu \left(z_l \,\overline l_L\gamma^\mu l_L
+ z_e\, \overline e_R\gamma^\mu e_R 
+ z_q \,\overline q_L\gamma^\mu q_L
+ z_u\, \overline u_R\gamma^\mu u_R
+ z_d\, \overline d_R\gamma^\mu d_R  \right) ~,
\ee
where $q_L=(u_L,d_L)$ and $l_L=(\nu_L,e_L)$ are $SU(2)_W$ doublets,
while $z_l$, $z_e$, $z_q$, $z_u$, and $z_d$ are the $U(1)_z$ charges
of the leptons and quarks.
The ensuing couplings at low energy of the $Z^\prime$ boson to electrons and 
nucleons are given by
\be
{\cal L}_{Z'}^g = Z'_\mu\left[
\overline{e} \gamma^\mu \left( g_V^e +  g_A^e \gamma_5 \right)e
+ \sum_{{\cal N} = n,p} \overline{\cal N} \gamma^\mu 
\left( g_V^{\cal N} + \gamma_5 g_A^{\cal N} \right) {\cal N} \right]~,
\label{vector-axial}
\ee
where the vector and axial couplings of the electron, proton and neutron 
are
\bear
&& g_{V,A}^e = \frac{g_z}{2} \left( z_e \pm z_l \right)~,
\nonumber \\ [2mm]
&& g_{V,A}^p = \frac{g_z}{2} \left( 2 z_u  + z_d \pm 3 z_q \right)~,
\nonumber \\ [2mm]
&& g_{V,A}^n = \frac{g_z}{2} \left( z_u  + 2 z_d \pm  3 z_q \right)~.
\label{vector-axial-couplings}
\eear
In addition to these dimension-4 interactions, there are higher-dimensional
interactions as in Eq.~(\ref{nucleon}), with 
$P_{\mu\nu} = \partial_\mu Z_\nu - \partial_\nu Z_\mu$,
describing magnetic- and electric-like dipole couplings.

The $U(1)_z$ charges may be treated as arbitrary real parameters. However, there 
are various requirements that any 
self-consistent theory that includes the $U(1)_z$ gauge group has to satisfy.
The $SU(3)_C\times SU(2)_W\times U(1)_Y\times U(1)_z$ gauge theory
must be anomaly free, so that the $U(1)_z$ charges must satisfy several 
cubic and linear equations. Furthermore, the quark and lepton charges are expected to be 
commensurate numbers ({\it i.e.}, their ratios are rational numbers),
which makes it much harder to satisfy the cubic equations.
It turns out \cite{Batra:2005rh}, however, that all anomaly cancellation conditions may be satisfied 
while keeping $z_l$, $z_e$, $z_q$, $z_u$, and $z_d$ arbitrary, provided there are enough 
additional fermions charged under $SU(3)_C\times SU(2)_W\times U(1)_Y\times U(1)_z$.
Those new fermions charged
under the standard model gauge group have not been seen in collider experiments so far, 
so that they must be heavier than a few hundred GeV. 
Given that those fermions must be chiral with respect to $U(1)_z$, their masses 
are less than $4\pi \langle \varphi \rangle$. Hence, the $U(1)_z$ breaking
VEV must be of the order of the electroweak scale, or larger, implying that 
$z_\varphi g_z \lae 10^{-14} - 10^{-31}$ for a $Z^\prime$-induced force
of range between a micrometer and the Earth-Sun distance.
Notice that this constraint may be satisfied even if
$g_z$ is of order one: $z_\varphi$ may be extremely small, and this situation
could arise naturally in theories involving
kinetic mixing of several $U(1)$ gauge groups \cite{Holdom:1985ag},
or gauge fields localized in extra dimensions \cite{Batell:2005wa}.

New fermions charged under the standard model gauge group may be avoided
in the case of ``nonexotic'' $Z^\prime$ (see Ref.~\cite{Appelquist:2002mw}), 
where the set of values for $z_l$, $z_e$, $z_q$, $z_u$, and $z_d$ is restricted 
such that 
\bear
&& g_{V,A}^p + g_{V,A}^e = 0 ~,
\nonumber \\ [2mm]
&& g_{V,A}^n = \frac{g_z}{2} \left[ \left(4\pm 3\right) z_q - z_u \right] ~.
\eear
Consequently, the long-range forces induced by nonexotic $Z^\prime$ exchange
are proportional to the number of neutrons.
For $z_q=z_u$ we recover the $U(1)_{B-L}$ gauge group
discussed at the beginning of Section 5.1; the associated $Z^\prime$ has no axial 
couplings, while its vector coupling to neutrons
is extremely constrained by tests of the material dependence of the inverse square law, 
$g_V^n = 3 z_q g_z \ll 10^{-19}$.
The particular case of $z_q=z_u=0$ corresponds to the paraphoton.

For $z_q \neq z_u$, even though in the case of nonexotic $Z^\prime$ the 
$U(1)_z$-breaking VEV is not required to induce a large mass for new fermions, 
a certain charge times the gauge coupling must still be very small.
To see this, note that the quark and charged-lepton mass terms have a 
$U(1)_z$ charge of $z_q - z_u$. If the Higgs doublet carries charge 
$z_q - z_u$, then the quark and lepton masses are generated as in the Standard Model,
but $z_H g_z$ must be very small such that the $Z^\prime$ is light enough
to mediate macroscopic forces.
If the Higgs doublet has zero $U(1)_z$ charge, then 
the masses of the up and down quarks,
and of the electron, should be generated by higher-dimensional operators, such as
\be
\lambda_d \frac{\varphi}{M}\overline{q}_L d_R H  ~~,
\ee 
where $\lambda_d$ is a dimensionless parameter smaller than $4\pi$ and $M$ is some 
mass scale larger than $\langle \varphi \rangle$. Therefore, the down-quark mass
requires a VEV $\langle \varphi \rangle$ in the MeV range or larger, so that
the range for the $Z^\prime$ mass $m_0$ considered here requires
$z_\varphi g_z \lae 10^{-9}$.

The only alternative to nonexotic $U(1)_z$ charges that would still avoid the presence of 
new fermions charged under the Standard Model gauge group involves  
generation-dependent $U(1)_z$ charges for the quarks and leptons.
For example, the electron contributions to the anomalies may be canceled by the muon ones
if the charges for the first- and second-generation leptons have opposite signs.
In this case $\langle \varphi \rangle$ may be much lower than in the case of nonexotic 
$U(1)_z$  with Higgs doublet charge different than $z_q-z_u$, 
but it still needs to be above $10^{-2}$ eV in order
to accommodate the solar neutrino oscillations.
We emphasize though that a low value for $\langle \varphi \rangle$
would in turn lead to the question of what 
stabilizes the hierarchy between $U(1)_z$ breaking scale and  the electroweak scale.

Despite the caveats discussed above, the various couplings of the ultra-light 
$Z^\prime$ may be treated in general as independent parameters.
It is interesting to observe that any of the vector or axial couplings 
of the electron, proton or neutron, given in Eq.~(\ref{vector-axial-couplings}),
may vanish even when the charges of the left- and right-handed quarks and leptons
are nonzero. That happens when the charges satisfy certain linear equations
(for example, $3 z_q = - z_u  - 2 z_d$ would imply that the neutron has 
no vector coupling to $Z'_\mu$), which may conceivably be consistent with 
some grand unified group. 

The amplitude for $Z^\prime$ exchange between an electron and a nucleon  
may be written as
\be
{\cal A}\!\left(\vec{q},\vec{P}\right) = -\frac{1}{\vec{q\,}^2}
\left(T^\nu-\frac{2 i v_h}{M^2} S^\nu \right)
\left(T'_\nu+\frac{2 i v_h}{M^2} S'_\nu \right)
\label{amplitude-z}
\ee
where $S^\nu$, defined in Eq.~(\ref{s-nu}), involves the effects of the 
magnetic- and electric-like dipole couplings of Eq.~(\ref{nucleon}), while
\be
T^\nu \equiv \overline{u}_e(P + q/2) \,\gamma^\mu \left(g_V+g_A\gamma_5
\right) u_e(P - q/2) ~,
\ee
involves the effects of the vector and axial couplings of 
Eq.~(\ref{vector-axial}).
In the nonrelativistic limit, the time-like component of
$T^\nu$ is 
\be
T^0=g_V^e\left [1-\frac{i \vec \sigma \cdot \left(\vec P\times
\vec q\right )}{4 m_e^2}\right]+g_A^e\,\frac{\vec \sigma\cdot\vec P}{m_e}~,
\ee
and the space-like component is
\be
\vec{T} = \frac{g_V^e}{2m_e}\left(2\vec P-i\, \vec
q\times\vec\sigma\right)+g_A^e\left
[\vec \sigma+\frac{i}{4m_e^2}\left(\vec P\times \vec q- 2i \vec
P\, \vec\sigma\cdot\vec P+\frac{i}{2} \vec q \,\vec\sigma\cdot\vec q
\right) \right] ~.
\ee
$T^{\prime 0}$ and $\vec{T}^{\, \prime}$ have analogous expressions, with
the electron couplings replaced by nucleon couplings,
$\vec\sigma$ replaced by $\vec\sigma'$, and a sign change for the
terms linear in momenta. 

We find that the majority of 
the long-range potentials listed in Eqs.~(\ref{yukawa-potential}), 
(\ref{sigma}), (\ref{static}) and (\ref{velocity}) 
may be induced by $Z^\prime$ exchange. Their momentum-independent 
coefficients can be derived by comparing Eqs.~(\ref{pot})
and (\ref{amplitude-z}).
As expected, there is a Yukawa potential between
unpolarized objects like in (\ref{yukawa-potential}) with a coefficient
\be
f_1^{e{\cal N}}(0,0) =  g_V^e \, g_V^{\cal N} ~.
\ee
All three potentials between a polarized object and an unpolarized
object, shown in Eq.~(\ref{sigma}), have nonzero coefficients: 
\be
f_r^{e{\cal N}} = - 4 g_V^{\cal N} \, \frac{v_h m_e}{M^2} {\rm Im} C_e ~,
\ee
for the monopole-dipole potential (this is a linear combination of ${\cal V}_9$ 
and ${\cal V}_{10}$), and 
\bear
f_\perp^{e{\cal N}} & = & \left( \frac{1}{2} + \frac{m_e}{m_{\cal N}} \right) 
g_V^e \, g_V^{\cal N} 
+ \frac{m_e^2}{2 m_{\cal N}^2} g_A^e\, g_A^{\cal N}
+ 4\left(1+\frac{m_e}{m_{\cal N}}\right)
g_V^{\cal N} \, \frac{v_h m_e}{M^2} {\rm Re} C_e  ~,
\nonumber \\ [4mm]
f_v^{e{\cal N}} & = & 2\left(1+\frac{m_e}{m_{\cal N}}\right) g_A^e \, g_V^{\cal N}
\label{onespin}
\eear
for the velocity-dependent potentials (these are linear combinations of 
${\cal V}_4$ and ${\cal V}_5$, and of ${\cal V}_{12}$ and ${\cal V}_{13}$, respectively).

In the case of two polarized bodies, all three static spin-spin potentials in 
Eq.~(\ref{static}) receive contributions with coefficients given by
\bear
f_{2}^{e{\cal N}}(0,0) & = & - g_A^e \, g_A^{\cal N}  ~,
\nonumber \\ [4mm]
f_{3}^{e{\cal N}}(0,0) & = & \frac{m_e}{4 m_{\cal N}} \, g_V^e \, g_V^{\cal N} 
+ \frac{1}{8}  \left(1+\frac{m_e^2}{m_{\cal N}^2}\right) g_A^e \, g_A^{\cal N} 
- \frac{v_h m_e}{M^2} \left ( g_V^e {\rm Re} C_{\cal N}
- \frac{m_e}{m_{\cal N}} g_V^{\cal N}{\rm Re} C_e  \right) 
\nonumber \\ [2mm]
&& \;+\; \left.f_{3}^{e{\cal N}}(0,0)\right|_{\gamma^\prime}  ~,
\nonumber \\ [3mm]
f_{11}^{e{\cal N}}(0,0) & = &\frac{1}{2} g_V^e\, g_A^{\cal N}+
\frac{m_e}{2 m_{\cal N}} g_A^e\,
g_V^{\cal N}- 2 \frac{v_h m_e}{M^2}\left( g_A^e {\rm Re} C_{\cal
N}-g_A^{\cal N} {\rm Re} C_e\right) ~~.
\label{spin-spin}
\eear
The velocity dependent spin-spin interactions in Eq.~(\ref{velocity})
also receive contributions, with coefficients:
\bear     
f_{6,7}^{e{\cal N}}(0,0) & = & 2 \left(1+\frac{m_e}{M_{\cal N}}\right) \frac{v_h m_e}{M^2} 
\left( g_A^e {\rm Im} C_{\cal N}\mp g_A^{\cal N} {\rm Im} C_e\right) ~,
\nonumber \\ [4mm]
f_{8}^{e{\cal N}}(0,0) & = & -\frac{1}{2}
\left( 1+\frac{m_e}{m_{\cal N}}+ \frac{m_e^2}{2 m_{\cal N}^2}\right) g_A^e \, g_A^{\cal N}  ~,
\nonumber \\ [4mm]
f_{15}^{e{\cal N}}(0,0) & = & \frac{v_h m_e}{M^2}\left[  
\left(\frac{1}{2}+\frac{m_e}{m_{\cal N}}\right) g_V^e {\rm Im} C_{\cal N} \
+ \frac{m_e}{m_{\cal N}} \left(1+\frac{m_e}{2 m_{\cal N}}\right) g_V^{\cal N} {\rm Im } C_e 
\right]  + \left.f_{15}^{e{\cal N}}(0,0)\right|_{\gamma^\prime} ~,
\nonumber \\ [4mm]
f_{16}^{e{\cal N}}(0,0) & = & \frac{m_e}{4 m_{\cal N}}\left(1+\frac{m_e}{m_{\cal N}}\right)
\left(g_V^e\, g_A^{\cal N}-g_A^e \, g_V^{\cal N}\right) 
+ \frac{v_h m_e}{M^2}\left [\left (\frac{1}{2}+\frac{m_e}{m_{\cal N}}+\frac{m_e^2}{m_{\cal
N}^2}\right) g_A^e {\rm Re} C_{\cal N} \right.
\nonumber \\ [1mm]
& & \left. + \, \left(1+\frac{m_e}{m_{\cal N}}+\frac{1}{2}\frac{m_e^2}{m_{\cal
N}^2}\right) g_A^{\cal N} {\rm Re} C_e\right]  ~~.
\eear
The last term in the above formulae for $f_{3}$ and $f_{15}$ represents the 
contribution from the magnetic- and electric-like dipole couplings,
given in Eq.~(\ref{paraphoton}).

The only operator from Eq.~(\ref{pot}) which does not contribute at
this order is ${\cal O}_{14}$. Once we include the higher-order corrections
proportional to additional powers of ${\vec q\,}^2$,  ${\cal O}_{14}$ is also
generated in the vector exchange. As previously mentioned, this
contribution is however suppressed by $m_0^2/m_e^2$, so it is too
small to be interesting in practice.

Let us now discuss the limits on the couplings of a low-mass $Z^\prime$.
The  limits on the Yukawa potential between 
unpolarized objects, shown in Eq.~(\ref{f1-limit}),
translate into a limit on the vector coupling of the neutron, and on the 
sum of the vector couplings of the electron and proton:
\be 
\left|g_V^n\right| \; , \; \left|g_V^e + g_V^p\right| \lae 10^{-20} ~,
\ee
for $1/m_0$ of order 1 cm, and almost four orders of magnitude stronger 
for $1/m_0 > 10^8$ m.

The experimental limits on the dipole-dipole potential ${\cal V}_3$
have been used in Section 5.1 to constrain the magnetic- 
and electric-like dipole couplings.
Once nonzero $U(1)_z$ charges are allowed, the
coefficient of ${\cal V}_3$ receives contributions that also depend on
the vector and axial couplings, as displayed in 
Eq.~(\ref{spin-spin}). The limit (\ref{f3-limit-ee}) becomes:
\be
- 0.8 \times 10^{-14} <
\frac{g_V^{e\; 2} + g_A^{e\; 2}}{16\pi^2\alpha}
- \frac{c_e^2 m_e^4}{\pi^2 \alpha M^4 }
< 3.2 \times 10^{-14} ~.
\ee
Barring accidental cancellations between the two terms above, 
we find
\be
\left|g_A^{e}\right| \; , \; \left|g_V^{e}\right| \lae 10^{-7} ~.
\ee
The limits from star cooling \cite{Raffelt:1999tx} are stronger by several 
orders of magnitude, but as discussed at the end of Section 5.1,
those limits may be avoided in the case of a spin-1 boson.
The indirect limit on the ${\cal V}_2$ potential derived in Eq.~(\ref{f2-limit})
provides the tightest restriction on the $Z^\prime$ axial coupling to the electrons:
\be
\left|g_A^{e}\right| \lae 10^{-17} ~.
\ee

The limits on the monopole-dipole potentials 
shown in Eqs.~(\ref{fr-limit}) and (\ref{fr-limit-n}) lead to the following constraints 
on the $Z^\prime$ couplings:
\bear
&& 4  c_e \, \frac{m_e^2}{M^2} \, \left| \sin\theta_e \; g_V^n\right| \lae 10^{-30} - 10^{-36} 
\;\; , \; \; {\rm for} \; 1/m_0 \gae \left(1 - 10^{11}\right)\,{\rm m}~~,
\nonumber \\ [3mm]
&& 4  c_n \, \frac{m_n^2}{M^2} \, \left| \sin\theta_n \; g_V^n\right| \lae 10^{-27} - 10^{-33} 
\;\; , \; \; {\rm for} \; 1/m_0 \gae \left(1 - 10^{6}\right)\,{\rm m}~~,
\eear 
There are also analogous constraints with $g_V^n$ replaced by $g_V^p + g_V^e$.

The new tests on velocity dependent potentials
of the type ${\cal V}_4+{\cal V}_{5}$ and ${\cal V}_{12}+{\cal V}_{13}$,
presented in Ref.~\cite{Heckel2006},
set preliminary limits
on the combination of couplings of the type $g_V^e$, $g_V^n$,
and ${\rm Re} \, c_e/M^2$ shown in Eq.~(\ref{onespin}).
In particular, the constraints on products of axial and vector 
couplings arising from the second 
Eq.~(\ref{fv-limit}) are extremely strong:
\be 
\left| g_A^e g_V^n \right| \; , \; 
\left| g_A^e \left(g_V^e + g_V^p\right) \right|  \lae 10^{-55} ~~,
\label{bestlimit}
\ee
for $1/m_0> 10^{11}$ m.

\section{Spin-0 exchange forces}
\setcounter{equation}{0}

A very light spin-0 particle, $\phi$, can have scalar and pseudoscalar couplings to
electrons and nucleons in the low-energy effective Lagrangian:
\be
{\cal L}_\phi =- \phi\, \overline{e} \left(g^e_{\rm S} + i\gamma_5g^e_{\rm P}\right) e 
- \phi\,\overline{\cal N} \left(g^{\cal N}_{\rm S} 
+ i\gamma_5g^{\cal N}_{\rm P}\right){\cal N} ~.
\label{scalar}
\ee
Any higher-dimensional coupling of $\phi$ to electrons or nucleons can be reduced
to the terms in Eq.~(\ref{scalar}) by integrating by parts and using the equations of
motion, so that they  do not give rise to new types of potentials.

The amplitude for electron-nucleon scattering due to the exchange of $\phi$
is given by Eq.~(\ref{pot}) with contributions from the operators
${\cal O}_1$ and  ${\cal O}_{4,5}$  for two scalar couplings,  
${\cal O}_3$ for two pseudoscalar couplings,
and ${\cal O}_{9,10}$ and ${\cal O}_{15}$ for
one scalar and one pseudoscalar coupling.
The only $f_i(0,0)$ coefficients that do not vanish are given by:
\bear
f_1^{e{\cal N}}(0,0) &=& -g^e_{\rm S} g^{\cal N}_{\rm S} 
\nonumber \\ [2mm]
f_3^{e{\cal N}}(0,0)&=& - \frac{m_e}{4 m_{\cal N}} g^e_{\rm P} g^{\cal N}_{\rm P}
\nonumber\\ [2mm]
f_{4,5}^{e{\cal N}}(0,0)&=&-\frac{1}{4}\left( 1 \pm \frac{m_e^2}{m_{\cal N}^2}\right)
g^e_{\rm S} g^{\cal N}_{\rm S} 
\nonumber \\ [2mm]
f_{9,10}^{e{\cal N}}(0,0)&=&\frac{1}{2}\left( g^e_{\rm P}g^{\cal N}_{\rm S}
\mp g^e_{\rm S}g^{\cal N}_{\rm P} \frac{m_e}{m_{\cal N}}\right)
\nonumber\\ [2mm]
f_{15}^{e{\cal N}}(0,0)&=& \frac{m_e}{4 m_{\cal N}}
\left(g^e_{\rm S}g^{\cal N}_{\rm P} - g^e_{\rm P}g^{\cal N}_{\rm S} 
 \frac{m_e}{m_{\cal N}}\right) ~~.
\label{f-scalars}
\eear

The spin-independent potential between two macroscopic objects induced 
by $\phi$ exchange is given in Eq.~(\ref{yukawa-potential}), with $f_1^{e\cal N}(0,0)$
dependent on the scalar couplings as shown in Eq.~(\ref{f-scalars}), and analogous 
expressions for the $f_1^{ee}(0,0)$ and $f_1^{\cal NN}(0,0)$ coefficients.
The limits from tests of the material dependence of the $1/r^2$ force
[see Eq.~(\ref{f1-limit})] give
\be 
\left|g_{\rm S}^n\right| \; , \; \left|g_{\rm S}^e + g_{\rm S}^p\right| 
\lae 10^{-20} - 10^{-24} ~,
\label{f1-limit-couplings}
\ee
depending on the range of the interaction, 
which is set by $1/m_0$ where $m_0$ is the $\phi$ mass.

The limit (\ref{f3-limit-ee}) on the coefficient of the 
dipole-dipole potential,  ${\cal V}_3$, between electrons 
may be written as 
\be
\frac{(g^e_{\rm P})^2}{16\pi^2 \alpha } < 0.8 \times 10^{-14} ~,
\ee
so that the constraint on the pseudoscalar coupling of the electron is 
\be
|g^e_{\rm P}| < 0.96 \times 10^{-7} ~.
\label{gep}
\ee
The limit (\ref{f3-limit-en}) on the coefficient of the 
dipole-dipole potential between electrons 
and neutrons gives $|g^e_{\rm P}g^n_{\rm P}| < 0.93 \times 10^{-7}$. 
Comparing with Eq.~(\ref{gep}), this 
places an almost irrelevant bound on the neutron pseudoscalar coupling,
$|g^n_{\rm P}| < 0.97$. Better limits (by three orders of magnitude)
on the pseudoscalar couplings to 
nucleons may be derived by considering 
two $\phi$ exchange \cite{Fischbach:1999iz}.

The potential between an 
unpolarized object and an object with polarized electrons is 
given by Eq.~(\ref{e-sigma}), with coefficients
\bear
f_r^{e\cal N} &=& - g_{\rm P}^e g_{\rm S}^{\cal N} ~,
\nonumber \\ [3mm]
f_\perp^{e\cal N} &=& \frac{1}{2} g_{\rm S}^e g_{\rm S}^{\cal N} ~,
\nonumber \\ [3mm]
f_v^{e\cal N} &=& 0 ~,
\eear
and analogous expressions for $f_r^{ee}$, $f_\perp^{ee}$ and $f_v^{ee}$.
The limits (\ref{fr-limit}) and (\ref{fr-limit-n}) on
the $f_r$ coefficients of the monopole-dipole potentials,  
based on the measurements presented
in Refs.~\cite{Heckel2006} and \cite{Venema1992}, respectively, yield constraints 
on products of scalar and pseudoscalar couplings:
\bear
&& \left| g_{\rm P}^e g_{\rm S}^n \right| \; , \; 
\left| g_{\rm P}^e \left(g_{\rm S}^e + g_{\rm S}^p \right)\right| 
\lae 10^{-30} - 10^{-36} 
\;\; , \; \; {\rm for} \; 1/m_0 \gae \left(1 - 10^{11}\right)\,{\rm m}~~,
\nonumber \\ [3mm]
&& \left| g_{\rm P}^n g_{\rm S}^n \right| \; , \; 
\left| g_{\rm P}^n \left(g_{\rm S}^e + g_{\rm S}^p \right)\right| 
\lae 10^{-27} - 10^{-33} 
\;\; , \; \; {\rm for} \; 1/m_0 \gae \left(1 - 10^{6}\right)\,{\rm m}~~.
\label{fr-limit-couplings}
\eear 
As discussed in Section 4, potentials of the type ${\cal V}_{4,5}$
have also been recently constrained in \cite{Heckel2006}. The limit 
(\ref{fv-limit}) provides a constraint on the scalar couplings different
than Eq.~(\ref{f1-limit-couplings}):
\be
\left| g_{\rm S}^e g_{\rm S}^n \right| \; , \; 
\left| g_{\rm S}^e \left(g_{\rm S}^e + g_{\rm S}^p \right)\right| 
\lae 10^{-31} ~~,
\label{fperp-limit-couplings}
\ee 
for $1/m_0 > 10^{11}$ m.

The star-cooling limit \cite{Eidelman:2004wy} on the pseudoscalar coupling
of the electron to a spin-0 particle, $|g^e_{\rm P}|< 10^{-12}$, is 
five orders of magnitude stronger than the one in Eq.~(\ref{gep}).
The scalar coupling to the electron is even more tightly constrained
by stellar dynamics, $|g^e_{\rm S}|< 10^{-14}$, which in conjunction with
the constraint from measurements of spin-independent long-range forces
given in Eq.~(\ref{f1-limit-couplings}) provides stronger limits than 
Eqs.~(\ref{fr-limit-couplings}) 
and (\ref{fperp-limit-couplings}).
In the case of the nucleons, the star-cooling limit is 
$|g^{\cal N}_{\rm P}| < 10^{-10}$.
Unlike the case of a spin-1 particle, where the astrophysical
constraints may be avoided as pointed out at the end of Section 5.1,
the star-cooling limits on spin-0 particles are quite robust 
(some attempts for relaxing the star-cooling constraint on the spin-0 coupling 
to photons are described in Ref.~\cite{Masso:2005ym}).

Furthermore, the constraints from searches for new long-range forces
may be relaxed in the case of forces mediated by spin-0 exchange 
if the new particle is self-interacting \cite{Gubser:2004uf}.
By contrast, the constraints on  new long-range forces
induced by spin-1 exchange are robust:
the self-interactions of the paraphoton are 
forbidden by the $U(1)$ gauge symmetry for operators of dimension 7 or less. 
Even in the case of
a $Z^\prime$, where the gauge symmetry is spontaneously broken,
self-interactions may be generated only by higher-dimensional 
operators which may be adequately suppressed.

\section{Conclusions}
\label{SectionConclusions}

Assuming energy and momentum conservation,
we have shown that rotational invariance restricts 
the long-range interaction between two fermions to a sum over 
16 spin-dependent potentials, given in center-of-mass frame by 
Eqs.~(\ref{even}) and (\ref{odd}).
The dependence of the potentials on the separation between the 
fermions, $r$, is shown in Eqs.~(\ref{yukawa-potential}), (\ref{sigma}),  
(\ref{static}) and (\ref{velocity})
for the case of one-boson exchange in a Lorentz invariant theory.
If the interaction is induced by 
two or more particles exchanged (as is the case for neutrino exchange 
\cite{Hsu:1992tg, Feinberg:1989ps}), 
or in the more exotic case where the kinetic term of the boson 
exchanged breaks Lorentz invariance (an example is given in
\cite{Arkani-Hamed:2004ar}), then different powers of $1/r$ appear in 
the potentials, as follows from Eqs.~(\ref{even}) and (\ref{odd}), 
but the spin dependence remains the same. 

Each of the 16 potentials 
has a dimensionless coefficient which is
momentum independent in the non-relativistic limit. 
The long-range forces between macroscopic objects depend on six different
two-particle potentials, $e^-e^-$, $pp$, $nn$, $e^-p$, $e^-n$ and $pn$,
each of them being described by a different set of 16 dimensionless
parameters (or only 12 parameters when the two fermions are identical).
Given that
searches for new macroscopic forces involve electrically-neutral objects,
the following set of parameters needs to be measured:
three coefficients of the potential between unpolarized bodies [see 
Eqs.~(\ref{yukawa-potential})], six coefficients for each of the 
three potentials between a polarized object and an unpolarized one [see 
Eqs.~(\ref{sigma})], six coefficients for each of the nine potentials
between two polarized objects given in Eqs.~(\ref{static}) and (\ref{velocity}),
with the exception of ${\cal V}_7$ where only three coefficients are nonzero.
The total number of parameters that need to be measured is 72.
Several of those have been already constrained, as discussed in Section 4.1.
Many others, both of the static and
velocity-dependent types, have not been explored in experiments so far.

In any quantum field theory that extends the Standard Model,
these 72 parameters are given in terms of the couplings of some 
very light particles. Therefore, one expects correlations between 
the various parameters. 
We have derived these correlations in the 
cases of one spin-0 or spin-1 particle exchange, in a general 
Lorentz-invariant theory. 
The constraints on the couplings of a spin-0 particle to
electrons and nucleons from measurements of 
spin-dependent macroscopic forces are weaker than the star-cooling
constraints. Moreover, they can be further relaxed in the presence of
self-interactions.
The opposite is true for spin-1 exchange, where the star-cooling constraints may 
be relaxed, while the searches for macroscopic forces are robust.

If an unbroken $U(1)$ symmetry is added to the Standard Model, 
theoretical motivation has led to considering only
magnetic- and electric-like dipole couplings of the new massless gauge boson 
to the electrons and nucleons.
As seen in section 5.1, this generates only two of the 16 long-range potentials
allowed by rotational invariance. The two potentials,
one static (${\cal V}_3$) and one velocity dependent (${\cal V}_{15}$), 
require both objects to be polarized in order to have an observable effect. 
If the new $U(1)$ symmetry is spontaneously broken,
the vector and axial couplings of the new low-mass gauge boson 
may also be present (see Section 5.2), 
and the result is that 15 of the 16 potentials are induced with 
unsuppressed coefficients,
while the remaining potential arises with an extra $(m_0/m_e)^2$ suppression,
where $m_0$ is the mass of the exchanged particle.
Remarkably, there are several spin-dependent potentials that 
fall off as $1/r$: ${\cal V}_2$, ${\cal V}_8$, ${\cal V}_{12} + {\cal V}_{13}$
and ${\cal V}_{14}$. Measurements of these
would be particularly sensitive to new one-boson exchanges in 
Lorentz invariant theories. 

Searching for the macroscopic interactions
discussed here could lead to the discovery of new
light particles, and at least would 
provide additional constraints on the properties of any new light particle
that couples to electrons or nucleons.

\bigskip\bigskip

{\bf Acknowledgments}: We would like to thank Eric Adelberger
for several insightful comments and stimulating discussions.

\bigskip\bigskip

\appendix{\bf \Large Appendix A: Vector identities}
\setcounter{equation}{0}
\renewcommand{\theequation}{A.\arabic{equation}}

In Section 2 we have stated that any scalar involving
two spins and two momenta may be written as a linear combination
of spin 16 operators. In this Appendix we give examples of 
such linear combinations.

In order to find relations between various operators, it is 
useful to note the following identities:
\bear
\left[\left(\vec a\times\vec b\right)\cdot\vec c \,\right]
\left[\left(\vec a\times\vec b\right)\cdot\vec d \,\right]
&=&
{\vec a\,}^2 {\vec b\,}^2 \left(\vec c\cdot\vec d\right) -
{\vec a\,}^2 \left(\vec b \cdot\vec c\right)\left(\vec b\cdot\vec d\right) -
{\vec b\,}^2 \left(\vec a \cdot\vec c\right)\left(\vec a\cdot\vec d \right)
\nonumber\\[2mm]
&+&\!\!\left(\vec a \cdot \vec b\right) \left[ \rule{0mm}{5mm} 
\left(\vec a\cdot\vec c\right)\left(\vec b\cdot\vec d\right) 
+ \left(\vec a\cdot\vec d\right)\left(\vec b\cdot\vec c\right) 
- \left(\vec a \cdot \vec b\right)\left( \vec c\cdot\vec d \right)\right] ~,
\nonumber
\eear
\bear
\left[(\vec a\times\vec b)\cdot\vec c \,\right]\left(\vec d\cdot\vec e\right) 
&=& \left[ (\vec a\times\vec b)\cdot\vec e \,\right] \left(\vec c\cdot\vec d\right) 
+ \left[ (\vec b\times\vec c)\cdot\vec e \,\right]\left(\vec a\cdot\vec d\right) 
- \left[(\vec a\times\vec c)\cdot\vec e \,\right]\left(\vec b\cdot\vec d\right) ~,
\nonumber\\[5mm]
\left(\vec a\times \vec b\right) \left(\vec c\times \vec d\right)
&=& \left(\vec a \cdot \vec c\right)\left( \vec b\cdot\vec d \right)
- \left(\vec a \cdot \vec d\right)\left( \vec b\cdot\vec c \right) ~,
\eear
where $\vec a$, $\vec b$, $\vec c$, $\vec d$ and $\vec e$ are arbitrary 
3-vectors.

Given that $\vec P \cdot \vec q = 0$, we find various nontrivial examples of
linear combinations: 
\bear
\left[ \vec \sigma \cdot \left(\vec P\times\vec q\right) \right]
\left[ \vec \sigma\,'\cdot \left(\vec P\times\vec q\right) \right] & = &
{\vec q\,}^2 {\vec P}^2 \CO_2 - m^2\left( {\vec P}^2 \CO_3 
+ {\vec q\,}^2 \CO_8 \right) ~,
\nonumber\\[3mm]
\left[\vec\sigma \cdot (\vec P\times\vec q) \right]
\left(\vec \sigma\,'\cdot\vec q\right) 
& = & m^3 \CO_{15} - \frac{m}{2}{\vec q\,}^2 \, \CO_{14} ~,
\nonumber\\[3mm]
i\left[\vec\sigma \cdot \left(\vec P\times\vec q\right) \right]
\left(\vec \sigma\,'\cdot\vec P\right) 
& = & m^3 \CO_{16} + \frac{m}{2} \vec P^2 \CO_{11} ~,
\nonumber\\[3mm]
\left(\vec\sigma \times \vec q\right) \left(\vec\sigma\,'\times\vec q\right) &=&
{\vec q\,}^2 \CO_2 - m^2 \CO_3 ~,
\nonumber\\[3mm]
\left(\vec\sigma \times \vec P\right) \left(\vec\sigma\,'\times\vec P\right) &=&
{\vec P}^2 \CO_2 - m^2 \CO_8 ~,
\nonumber\\[3mm]
i\left[\vec\sigma \times \left(\vec P\times\vec q\right) \right]
\cdot \vec\sigma\,' &=& - 2 m^2 \CO_7 ~,
\eear
where $\CO_i$ are the operators listed in Eqs.~(\ref{ops}) and (\ref{ops-prime}).
Based on these and other similar identities, one can prove by exhaustion
that the set of 16  operators $\CO_i$ is complete. 

\bigskip\bigskip
\appendix{\bf \Large Appendix B: Fourier transforms}
\setcounter{equation}{0}
\renewcommand{\theequation}{B.\arabic{equation}}

The Fourier transforms necessary  for obtaining the 
potentials induced by the exchange of one boson with normal
propagator (see Section 4.1) are given by 
\bear
\int \frac{d^3q}{(2\pi)^3} e^{i\vec q\cdot \vec r}
\frac{1}{{\vec q\,}^2 + m_0^2} ({\vec q\,}^2)^l
&= &\frac{1}{4 \pi r}  e^{-m_0 r} m_0^{2 l} (-1)^l ~~,
\nonumber\\[4mm]
\int \frac{d^3q}{(2\pi)^3} e^{i\vec q\cdot \vec r}
\frac{\vec{q}}
{{\vec q\,}^2 + m_0^2} ({\vec q\,}^2)^l
&= &\frac{i}{4 \pi r^2}  (1+m_0 r) e^{-m_0 r} m_0^{2 l} (-1)^l
\, \hat{\vec r}~~,
\nonumber\\[4mm]
\int \frac{d^3q}{(2\pi)^3} e^{i\vec q\cdot \vec r}
\frac{ \left(\vec{q}\cdot \vec{n}_1\right) 
\left(\vec{q}\cdot \vec{n}_2\right)}{{\vec q\,}^2 + m_0^2} ({\vec q\,}^2)^l
&= &-\frac{1}{4 \pi r^3}  e^{-m_0 r} m_0^{2 l} (-1)^l
\left[\rule{0mm}{5mm}(1+m_0r) \vec{n}_1\cdot \vec{n}_2\right.
\nonumber\\[2mm]
&-& \!\!\left.  \left(3+3 m_0 r+m_0^2r^2\right) \left(\hat{\vec r}\cdot \vec{n}_1\right) 
\left(\hat{\vec r}\cdot \vec{n}_2\right) \right] ~~,
\label{mass-integrals}
\eear
where $l\ge 0$ is an integer, while $\vec n_1$ and $\vec n_2$ are arbitrary 3-vectors.
For $l=1$ we obtain the results used in Section \ref{standard}, with 
$\vec n_1$ and $\vec n_2$ replaced by $\vec \sigma$, $\vec \sigma'$, $\vec v$ or 
combinations thereof, as needed for the various operators.

We also give here the Fourier transforms relevant for 
the cases where there are non-standard dispersion relations (see Section 4.2):
\bear
\int \frac{d^3q}{(2\pi)^3} \frac{e^{i\vec q\cdot \vec
r}}{({\vec q\,}^2)^k}
&=& 
\frac{1}{2\pi^2} \,r^{2k-3} \sin(k\pi)\Gamma(2-2k)
\nonumber\\[2mm]
&=&
- 
\frac{1}{[2 (k-1)]! \, 4\pi} \, r^{2k-3} ~~,  
\nonumber\\[5mm]
\int \frac{d^3q}{(2\pi)^3} \frac{e^{i\vec q\cdot \vec r}}{({\vec q\,}^2)^k} \vec{q}
&=& -i \vec \nabla\int \frac{d^3q}{(2\pi)^3} 
\frac{e^{i\vec q\cdot \vec r}}{({\vec q\,}^2)^k}
\nonumber\\[2mm]
&=& \frac{i(2k-3)}{[2 (k-1)]!\, 4\pi} \, r^{2k-4}\, \hat{\vec r} ~~,
\nonumber\\[5mm]
\int \frac{d^3q}{(2\pi)^3} \frac{e^{i\vec q\cdot \vec
r}}{({\vec q\,}^2)^k} \left(\vec q\cdot \vec n_1\right) \left(\vec q\cdot \vec n_2\right)
&=& \frac{(2k-3)}{[2 (k-1)]!\, 4\pi} \, r^{2k-5} 
\left[\hat{\vec n_1}\cdot\hat{\vec
n_2}+(2k-5) \left(\hat{\vec{r}}\cdot \vec n_1\right) \left(\hat{\vec r}\cdot \vec n_2\right) \right] ~,
\nonumber\\
\label{fourier-ghost}
\eear
where $k\ge 1$ is an integer. 

\bigskip


\vfil
\end{document}